\documentclass[11pt]{article}

\usepackage{amsfonts,amsmath,amssymb}
\usepackage{mathrsfs}
\usepackage{natbib}
\usepackage{graphicx}
\usepackage{multirow}
\usepackage{epsfig}
\usepackage{subfigure}
\usepackage{booktabs}
\usepackage{array}
\usepackage{tabularx}
\usepackage{slashed}
\usepackage{physics}
\usepackage{braket}
\usepackage{bm}
\usepackage{diagbox}
\usepackage{cancel}
\usepackage{mathtools}
\usepackage[usenames]{color}
\usepackage[dvipsnames]{xcolor}
\usepackage[utf8]{inputenc}
\usepackage{fontspec}
\usepackage{xeCJK}
\usepackage{hyperref}
\hypersetup{
  colorlinks=true,        
  linkcolor=ForestGreen,    
}

\newcommand{\mhi}{M_{\text{hi}}}

\newcommand{\chn}[3]{{{}^{#1}\!{#2}_{#3}}}
\newcommand{\cs}[2]{\chn{#1}{S}{#2}}
\newcommand{\chp}[2]{\chn{#1}{P}{#2}}
\newcommand{\cd}[2]{\chn{#1}{D}{#2}}
\newcommand{\cf}[2]{\chn{#1}{F}{#2}}
\newcommand{\cg}[2]{{}^{#1}{G}_{#2}}
\newcommand{\csd}{{\cs{3}{1}-\cd{3}{1}}}
\newcommand{\cpf}{{\chp{3}{2}-\cf{3}{2}}}
\newcommand{\cdg}{{\cd{3}{3}-\cg{3}{3}}}

\newcommand{\NNLO}{N$^2$LO}

\newcommand{\NdKet}[1]{\ket{\phi^{(#1)}}}

\newcommand{\ii}{\mathrm{i}}

\graphicspath{{FigsInPaper/}}
\begin{document}

\title{Perturbative calculations of nucleon-deuteron elastic scattering in chiral effective field theory}

\author{
Lin Zuo (左林)\\
College of Physics, Sichuan University, Chengdu 610065, China
\and
Wendi Chen (陈文棣)\\
Institute of Applied Physics and Computational Mathematics, \\
Beijing 100094, China
\and
Dan-Yang Pang (庞丹阳)\\
School of Physics, Beihang University, Beijing 100191, China
\and
Bingwei Long (龙炳蔚)\thanks{Email: bingwei@scu.edu.cn}\\
College of Physics, Sichuan University, Chengdu 610065, China\\
Southern Center for Nuclear-Science Theory (SCNT),\\
Institute of Modern Physics, Chinese Academy of Sciences,\\
Huizhou 516000, Guangdong, China
}

\date{April 3, 2026}

\maketitle

\begin{abstract}
We develop a framework for calculating nucleon-deuteron scattering using the Faddeev equations, employing strict perturbation theory to treat subleading interactions in chiral effective field theory (ChEFT).
Rather than evaluating the distorted-wave expansion directly, our approach solves a hierarchy of integral equations to obtain subleading scattering amplitudes.
We benchmark the method against the wave-packet continuum discretization.
This framework benefits from the fact that renormalization-group-invariant chiral forces involve only a limited number of two-body partial waves at leading order.
We use it to calculate differential cross sections and analyzing powers for nucleon-deuteron elastic scattering up to next-to-leading order.
\end{abstract}

\section{Introduction}

The nucleon-deuteron ($Nd$) system provides an important testing ground for chiral nuclear forces.
Not only does it test the prediction of chiral nucleon-nucleon ($NN$) potentials for the three-nucleon ($3N$) system~\cite{witala_cornelius_gloeckle_1988, Glockle:1996jg, Huber:1995zza, Gloeckle:1990bi, Witala:1999sg, Kievsky:1997bd, Witaa:2003en}, but it also helps in understanding the role of $3N$ potentials by quantifying their importance via power counting~\cite{Girlanda:2023znc, Witala:2000am, Miller:2022beg, Margaryan:2015rzg, Golak:2014ksa, Witala:2013ioa, Epelbaum:2019zqc, Witala:2022rzl}.
While the triton bound state and its properties have been the natural choice, the scope of investigation is limited by the quantum numbers of the triton, e.g., the total angular momentum $J = \frac{1}{2}$.
Nucleon-deuteron scattering, especially neutron-deuteron ($nd$) scattering, offers more probes of the $3N$ system without having to account for precision-level details of electromagnetic or weak interactions.
In recent years, strict perturbative treatment of subleading-order interactions has been increasingly advocated in the development of effective field theories (EFTs) for nuclear physics. 
Perturbation theory on top of a nonperturbative leading order (LO) has the advantage of disentangling subleading interactions from the LO ones.
For instance, two-pion exchange (TPE) potentials in $NN$, although subleading, can become much stronger than the one-pion exchange (OPE) potential at intermediate momenta near the ultraviolet cutoff $\sim \Lambda$. Perturbative calculations make it clear how TPEs are subtracted in the ultraviolet region by subleading contact interactions, thereby producing small corrections to on-shell amplitudes~\cite{Long:2011qx, Long:2011xw, Long:2012ve, PavonValderrama:2011fcz}.
However, a perturbative treatment complicates the computation when the $3N$ continuum problem is already more involved than the bound-state problem.
The main focus of this paper is to develop a technique to perform these perturbative calculations for chiral nuclear forces in the context of nucleon-deuteron scattering.

For the nuclear force, we use the power counting developed in Refs.~\cite{Long:2011xw, Long:2012ve}, and later modified by Ref.~\cite{Wu:2018lai}, to organize chiral $NN$ forces at different orders.
More specifically, we follow Refs.~\cite{Long:2011xw, Long:2012ve} to treat the so-called nonperturbative-pion channels: $\cs{1}{0}$, $\csd$, and $\chp{3}{0}$.
For other partial waves where OPE is considered as a perturbation, the power counting laid out in Ref.~\cite{Wu:2018lai} is adopted.
This scheme was explained recently in Ref.~\cite{Andis:2025fsg}, and the part relevant to this paper is reviewed in more detail in Sec.~\ref{sec:PertNLO}.
The same power counting is also employed to study electroweak processes in Refs.~\cite{Shi:2022blm, Liu:2022cfd, Andis:2025fsg}.
We note that the power counting of Refs.~\cite{Long:2011xw, Long:2012ve} has been examined with Bayesian analyses for $NN$ scattering data~\cite{Thim:2023fnl, Thim:2024yks} and has been used to study the structure of the deuteron and triton~\cite{Thim:2025vhe}.
In constructing the power counting for two-body potentials, these works adopt renormalization-group (RG) invariance as a guideline, which requires that the phase shifts be independent of the momentum cutoff, an arbitrary parameter of the ultraviolet regularization.

The OPE potential is the most important long-range nucleon force in chiral EFT, and it has a tensor component that behaves like $1/r^3$ at short distances.
For attractive singular potentials, such as the OPE tensor force in $\csd$, $\chp{3}{0}$, and $\cpf$, RG invariance requires a contact potential, often referred to as a counterterm, to appear at LO if OPE is considered nonperturbative in that partial wave~\cite{Beane:2000wh, PavonValderrama:2004nb, Nogga:2005hy, Long:2007vp}, even though the naive dimensional analysis (NDA) adopted by Weinberg's power counting would not require one~\cite{Weinberg:1990rz, Weinberg:1991um, Weinberg:1992yk}.
Because there are, in principle, an infinite number of attractive singular channels for OPE, one would have to invoke an infinite number of counterterms already at LO.
This conundrum is avoided once we recognize that OPE does not need to be resummed nonperturbatively in the Lippmann-Schwinger or Schr\"odinger equation for sufficiently high orbital angular momentum~\cite{Wu:2018lai, Kaplan:2019znu, Birse:2005um} and that NDA is restored if OPE is treated in pure perturbation theory.
We follow Ref.~\cite{Wu:2018lai} in letting OPE enter at LO only in $\cs{1}{0}$, $\csd$, and $\chp{3}{0}$ and at next-to-leading order (NLO) in all other waves.
Not only does this development of two-body chiral forces serve as the foundation of our study of nucleon-deuteron scattering, but it also illustrates the intertwined logic among renormalization, power counting, and perturbation theory for subleading interactions.

A universal feature of EFTs is the increasing momentum power of higher-order interactions.
Although this facilitates expansions for low-momentum initial and final states where momenta $Q$ are well below the breakdown scale $\mhi$, these higher-order interactions are not necessarily small for intermediate states with momenta up to the ultraviolet cutoff $\Lambda \gtrsim \mhi$.
Perturbative renormalization of subleading orders has been advocated as a reliable way to ensure that the resulting large contributions from intermediate states can be absorbed into low-energy constants (LECs).
Numerous applications of strict perturbation theory in pionless and chiral EFTs can be found in the recent review in Ref.~\cite{Hammer:2019poc}.

A key aspect of the power counting of chiral forces used in this paper is that the LO potentials are nonzero only in a limited number of $NN$ partial waves.
In the distorted-wave expansion, perturbation theory for subleading potentials is applied by directly evaluating matrix elements between the LO asymptotic states.
By contrast, our technique solves a hierarchy of integral equations at subleading orders, all of which share the kernel from the LO equation but have a distinct driving term at each order.
This approach to implementing perturbation theory for subleading-order interactions is in line with the methods developed in Refs.~\cite{Vanasse:2013sda, Konig:2019xxk} for pionless-EFT calculations of few-body systems.
A similar framework has been developed in Ref.~\cite{Andis:2025fsg} to calculate the longitudinal response function of the deuteron perturbatively.

This paper is organized as follows: in Sec.~\ref{sec:Framework}, we describe the numerical framework for solving the Faddeev equation, focusing on the contour-deformation method.
In Sec.~\ref{sec:PertNLO}, we give details of the perturbative treatment of the NLO potentials. We then present benchmark calculations to validate our methods in Sec.~\ref{sec:Bench}. The LO and NLO results for $Nd$ elastic scattering are presented and discussed in Sec.~\ref{sec:Results}, and we conclude with a summary in Sec.~\ref{sec:summary}.

\section{Faddeev equation with Deformed Contour\label{sec:Framework}}

\subsection{Jacobi partial-wave basis}
In our calculations, we expand the Faddeev equation in the Jacobi partial-wave basis~\cite{Glockle-Fewbody83}:
\begin{equation}
|p q \alpha \rangle \equiv | pq\; (ls)j \left (\lambda \frac{1}{2} \right ) I J \left (t \frac{1}{2} \right) T \rangle \, .\label{eqn:Jacobi3NBasis}
\end{equation}
Here, $p$ and $q$ are the magnitudes of the Jacobi momenta: $\vec{p}$ is the relative momentum of the subsystem (nucleons 1 and 2), and $\vec{q}$ is the momentum of the ``spectator'' (nucleon 3) relative to the center of mass of the subsystem. The quantum numbers $l$, $s$, $j$, and $t$ are, respectively, the orbital angular momentum, spin, total angular momentum, and isospin of the subsystem; $\lambda$ and $I$ are the orbital angular momentum and total angular momentum of the spectator; and $J$ and $T$ are the total angular momentum and total isospin of the $3N$ system. The Jacobi partial-wave basis is partially antisymmetrized for the $NN$ subsystem, i.e., $l+s+t =$ odd, and the parity is given by $P=(-1)^{l+\lambda}$. This paper focuses on $Nd$ elastic scattering, for which $T=\frac{1}{2}$. For simplicity, we use the collective label $\alpha$ to denote these discrete quantum numbers. Our choice for the normalization of the Jacobi partial-wave basis is as follows:
\begin{align}
\braket{p'q'\alpha'|pq\alpha}=\delta_{\alpha'\alpha}\dfrac{\delta(p'-p)}{p'p}\dfrac{\delta(q'-q)}{q'q}\, .
\label{eqn:JacobiNmlz}
\end{align}

We require the initial and final wave functions, $\ket{\phi}$, describing configurations in which the nucleon and the deuteron are far apart. To project an initial or final wave function onto the partial-wave basis, we first enumerate all $3N$ channels that include a deuteron $NN$ channel:
\begin{equation}
  | \alpha_d \rangle \equiv  | (l_d 1)1 \left (\lambda \frac{1}{2} \right) I J \left (t \frac{1}{2} \right ) \dfrac{1}{2} \rangle \, ,
\end{equation}
where the orbital angular momentum is restricted to $0$ or $2$ (i.e., $l_d = 0$ or $2$).
For a given value of the total $3N$ angular momentum $J$, there may be multiple combinations of $\lambda$ and $I$ that yield a valid $\alpha_d$ channel.
The initial $Nd$ state, with definite $J$, $\lambda$, and $I$, is then constructed as
\begin{equation}
  |\phi_{\lambda I}^{J}\, q_0 \rangle = \sum_{l_d = 0, 2} \int dp p^2 \varphi_{l_d}(p)|p q_0 \alpha_d\rangle \, ,
  \label{eqn:NdInitialState}
\end{equation}
where $\varphi_{l_d}(p)$ denotes the deuteron wave function in the $NN$ partial wave $l_d$, and $q_0$ is the center-of-mass momentum of the incoming nucleon.

However, it is more customary to use the $\lambda \Sigma$ basis for defining the phase shifts and mixing angles in $Nd$ elastic scattering, where $\Sigma$ is the channel spin, i.e., the total spin of the deuteron and the incoming or outgoing nucleon~\cite{Seyler:1969sii}: 
\begin{align}
\vec{\Sigma}=\vec{s}_d+\vec{s}_N\, .
\end{align}
The $\lambda\Sigma$ and $\lambda I$ bases are related as follows:
\begin{align}
    \ket{\phi_{\lambda\Sigma}^J\, q_0}=\sum_I (-1)^{J-I}\sqrt{I\Sigma}\left \{\begin{array}{ccc}
     \lambda &\frac{1}{2}&I\\[0.5ex]
        s_d&J&\Sigma
    \end{array}
    \right \}\ket{\phi_{\lambda I}^J \, q_0}\, .
    \label{eqn:SigmaITrans}
\end{align}

For nucleon-deuteron scattering, the Faddeev equation can be greatly simplified because the three nucleons are identical fermions.
Particle-exchange operators are crucial for enforcing the fermionic nature of the nucleons. If $P_{ij}$ denotes the exchange of the two nucleons labeled $i$ and $j$, the $3N$ permutation operator $P$ is the sum of the cyclic and anticyclic permutations of the three nucleons \cite{Glockle-Fewbody83}:
\begin{equation}
    P \equiv P_{12}P_{23} + P_{13}P_{23} \, .
\end{equation} 
In practice, the permutation operator $P$ is projected onto the Jacobi partial-wave basis \eqref{eqn:Jacobi3NBasis}:
\begin{align}
    &\bra{p'q' \alpha'}P\ket{pq \alpha} 
    = \delta_{J' J} \delta_{T' T}  \int_{-1}^1 dx \dfrac{\delta(p'-\pi_1)}{p'^{l'+2}} \dfrac{\delta(p-\pi_2)}{p^{l+2}} G_{\alpha' \alpha}(q'qx) \, , \label{eqn:Rep_PMatrix}
\end{align}
where
\begin{align}
    \pi_1(q',q)&=\sqrt{q^2+\frac{1}{4}q'^2+xq'q}\, ,\\
    \pi_2(q',q)&=\sqrt{\frac{1}{4}q^2+q'^2+xq'q}\, ,\label{eqn:ShiftMmta_Gqqx}
\end{align}
and
\begin{align}
    G_{\alpha' \alpha}(q'qx)=\sum_k P_k(x)\sum_{\substack{l_1'+l_2'=l'\\l_1+l_2=l}}q'^{l_2'+l_2}q^{l_1'+l_1}g_{\alpha'\alpha}^{kl_1'l_2'l_1l_2}\, .
\end{align}
Here, $P_k$ denotes the Legendre polynomial of order $k$; $l_{1, 2}$ and $l_{1, 2}'$ are intermediate angular momenta that are summed over; the final-state quantum numbers, such as $l', \lambda', s', \cdots$, are collectively denoted by $\alpha'$; and the coefficient $g_{\alpha'\alpha}^{kl_1'l_2'l_1l_2}$ is defined as follows \cite{Glockle:1996jg}:
\begin{align}
g_{\alpha'\alpha}^{kl_1'l_2'l_1l_2} &=\left( - \right)\sqrt{\hat{l}\hat{s}\hat{j}\hat{t}\hat{\lambda}\hat{I}\hat{l}'\hat{s}'\hat{j}'\hat{t}'\hat{\lambda}'\hat{I}'}  
\left\{ \begin{array}{ccc}
\frac{1}{2} & \frac{1}{2} & t' \\
\frac{1}{2} & T & t
\end{array} \right\}\sum_{LS} \hat{L} \hat{S} 
\left\{ \begin{array}{ccc}
\frac{1}{2} & \frac{1}{2} & s' \\
\frac{1}{2} & S & s
\end{array} \right\} \nonumber\\
&\quad \times\left\{ \begin{array}{ccc}
l & s & j \\
\lambda & \frac{1}{2} & I \\
L & S & J
\end{array} \right\} 
\left\{ \begin{array}{ccc}
l' & s' & j' \\
\lambda' & \frac{1}{2} & I' \\
L & S & J
\end{array} \right\}  \hat{k} \left( \frac{1}{2} \right)^{l'_2 + l_1} 
\sqrt{\frac{(2l+1)!}{(2l_1)!(2l_2)!}} \nonumber\\
&\quad \times\sqrt{\frac{(2l'+1)!}{(2l'_1)!(2l'_2)!}} \sum_{ff'} 
\left\{ \begin{array}{ccc}
l_1' & l_2'&l' \\
\lambda' & L&f'
\end{array} \right\} 
\left\{ \begin{array}{ccc}
l_2 & l_1&l \\
\lambda & L&f
\end{array} \right\} 
C_{l'_2 \lambda' f'}^ {0\,  0\,  0} \ 
C_{l_1 \lambda f}^{0 \, 0 \, 0} \nonumber \\
&\quad \times \left\{ \begin{array}{ccc}
f' & l'_1&L \\
f & l_2&k
\end{array} \right\} 
C_{k l'_1 f}^{ 0 \, 0 \, 0} \ 
C_{k l_2 f'}^{ 0 \, 0 \, 0}\, ,
\end{align}
where $\hat{X}\equiv{2X+1}$; $L$ and $S$ denote, respectively, the total orbital angular momentum and the total spin allowed by the total angular momentum $J$; and $f$ and $f'$ are, again, the intermediate angular momenta used to facilitate recoupling.

\subsection{Inhomogeneous Faddeev equation} 

\begin{figure}
    \centering
    \includegraphics[scale=0.5]{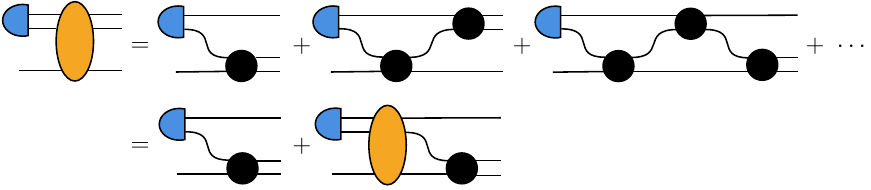}
\caption{Diagrammatic representation of the Faddeev equation. Solid lines represent nucleons; the solid circle denotes the two-body off-shell $t$-matrix; the blue half-circle denotes the deuteron; and the yellow blob denotes the Faddeev breakup amplitude $T$. Not all particle-exchange topologies are shown.
}
\label{fig:FeyDagm_fdvNoNdState}
\end{figure}

The inhomogeneous Faddeev equation is diagrammatically illustrated in Fig.~\ref{fig:FeyDagm_fdvNoNdState}.
We use chiral potentials in this study, and the $3N$ forces do not contribute up to NLO for renormalization purposes, as shown in Ref.~\cite{Song:2016ale} and verified in this work.
Therefore, only two-body potentials are considered here.
The yellow blob denotes the breakup amplitude $T$, which starts from an $Nd$ initial state $\phi$ and ends with three free nucleons in the final state.
The solid circle represents the full off-shell two-body $t$-matrix, which satisfies the Lippmann-Schwinger equation (LSE)
\begin{align}
    t&=V_2+V_2G_0t\, ,\label{eqn:LSE}
\end{align}
where $V_2$ is the two-body potential and $G_0$ is the free propagator.
The propagation of the breakup process, denoted by $T\ket{\phi}$, is given symbolically by the following equation~\cite{Glockle-Fewbody83, Glockle:1996jg}:
\begin{equation}
T\ket{\phi} = t P\ket{\phi} + t P G_0 T\ket{\phi} \, .
\label{eqn:Faddeev}
\end{equation}
The total energy of the $3N$ system, $E_3$, is related to the center-of-mass momentum $q_0$ of the incoming nucleon by:
\begin{equation}
    E_3= E_d+\dfrac{3q_0^2}{4m_N}\, ,\label{eqn:E3TotalEnergy}
\end{equation}
where $E_d = - B_d$ denotes the (negative) deuteron binding energy and $m_N$ is the nucleon mass.

The Faddeev equation \eqref{eqn:Faddeev} is solved in the $\lambda I$ basis to obtain the Faddeev breakup amplitude $\bra{pq\alpha} T\ket{\phi_{\lambda I}^J}$.
In turn, the $Nd$ elastic amplitude $U$ is computed from the following relation \cite{Glockle:1996jg}:
\begin{equation}
U^J_{\lambda'\Sigma', \lambda \Sigma}(q_0) = \braket{\phi'^J_{\lambda' \Sigma'}| PG_0^{-1} + PT | \phi^J_{\lambda \Sigma}}\, .
\label{eqn:ElasScatU}
\end{equation}
The basis transformation from $\lambda I$ to $\lambda \Sigma$ is carried out according to Eq.~\eqref{eqn:SigmaITrans}.
The $S$-matrix for elastic scattering is directly related to $U^J_{\lambda'\Sigma', \lambda \Sigma}$ by
\begin{align}
    S^J_{\lambda'\Sigma',\lambda \Sigma}(q_0)=\delta_{\lambda'\lambda}\delta_{\Sigma'\Sigma}-\ii\dfrac{4\pi}{3}q_0m_Ni^{\lambda'-\lambda}U_{\lambda'\Sigma',\lambda\Sigma}^J\, .\label{eqn:FromUMatToSMtat}
\end{align}
We follow Ref.~\cite{Seyler:1969sii} in parameterizing the $S$-matrix, thereby defining the phase shifts and mixing angles.

\subsection{Deformed Contour}

\begin{figure}[thb]
    \centering
    \includegraphics[scale=1]{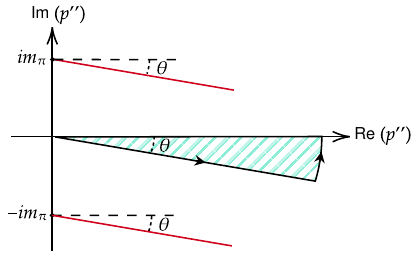}
    \caption{
    Diagram illustrating the analytic structure of $V^{1\pi}(p',p'')$ when the integration contours in both $p'$ and $p''$ are rotated by an angle $\theta$. The red lines trace the trajectories of the two branch points at $p'' = p' \pm i m_{\pi}$, which arise from the endpoint singularities of the integral.
    }
    \label{fig:PionPole}
\end{figure}

Projecting the abstract operator equations~\eqref{eqn:Faddeev}, \eqref{eqn:LSE}, and~\eqref{eqn:ElasScatU} onto the partial-wave basis yields integral equations for $NN$ and $3N$ dynamics.
The momentum-space integrals in these equations typically involve singularities.
In numerical computations, we employ contour deformation to circumvent these singularities.
Suppose $p''$ is the integration variable and that the original contour runs from $0$ to $\infty$ along the positive real axis; we deform the contour by rotating it clockwise by a small angle $\theta$:
\begin{equation}
    p'' = e^{-\ii\theta} x \, ,\; x \in \mathbb{R}^+ \, .
    \label{eqn:RotatedVar}
\end{equation}
In Fig.~\ref{fig:PionPole}, the rotated contour is illustrated by a ray at angle $\theta$.
For sufficiently large $p''$, the deformed contour returns counterclockwise to $+\infty$ on the real axis.
In practical calculations, we ensure that the integrand decays rapidly enough that the integral along the arc can be neglected.
This technique is akin to the ``complex scaling'' method used in many coordinate-space and momentum-space calculations of scattering and reaction processes, e.g., in Refs.~\cite{Ho:1983lwa, Moiseyev:1998gjp, Myo:2014ypa, Myo:2020rni,Hetherington:1965zza,Aaron:1966zz}.

The implementation begins with the LSE~\eqref{eqn:LSE}:
\begin{align}
t_{l'l}(p',p,E_2)=V_{l'l}(p',p) +\sum_{\Bar{l}'}\int \dd p''p''^2V_{l'\Bar{l}'}(p',p'')
 \dfrac{t_{\Bar{l}'l}(p'',p,E_2)}{E_2-\dfrac{p''^2}{m_N}+\ii\epsilon}\, , 
 \label{eqn:LSEOnJacobi}
\end{align}
where $p$ ($p'$) denotes the incoming (outgoing) relative momentum, and $E_2$ is the center-of-mass energy of the $NN$ pair.
We introduce the notation for coupled $NN$ partial waves: $\Bar{l} = l$ for uncoupled channels, while for coupled channels $\Bar{l}$ takes the values $j-1$ and $j+1$.
Not only is the $p''$ contour deformed by rotation, but the arguments of the full off-shell $t$-matrix $t(p', p, E_2)$ --- $p'$ and $p$ --- also lie along the rotated axis.
The most prominent singularity is the pole of the free $NN$ propagator at $p'' = \sqrt{m_NE_2}$. The deformed contour clearly avoids it.
In our numerical integrations, these ``rotated'' complex variables are represented by real Gauss--Legendre mesh points multiplied by a complex phase.
\begin{equation}
    p''_n = e^{-\ii \theta} x_n\, .
    \label{eqn:RotMesh}
\end{equation}
Thus, the integral is approximated by a sum: 
\begin{equation}
    \int dp'' f(p'') \approx e^{-\ii \theta} \sum_n w_n f(e^{-\ii \theta} x_n)\, ,
\end{equation}
where ${x_n}$ and ${w_n}$ are the standard abscissae and weights on the positive real axis.

In the ChEFT construction of $V_{l'\Bar{l}'}(p', p'')$, the contact potentials are usually polynomials in $p'$ and $p''$; therefore, they do not introduce any singularities in the contour integration. However, the pion-exchange components of $V_{l'\Bar{l}'}(p', p'')$ exhibit nontrivial singularities. The OPE potential $V^{1\pi}$ admits the following integral representation in momentum space:
\begin{equation}
    V^{1\pi}_{l'\Bar{l}'}(p', p'')\propto \int_{-1}^1 \dd x \dfrac{P_{l'}(x)}{p'^2+p''^2-2xp'p''+m_{\pi}^2}\, ,\label{eqn:OffShell_PionPole}
\end{equation}
where $m_{\pi}$ is the pion mass.
For fixed $p'$, $V^{1\pi}_{l'\Bar{l}'}(p', p'')$ has branch points in the complex $p''$-plane at $\pm p' \pm \textrm{i} m_\pi$, arising from the endpoint singularities of the integral, as illustrated in Fig.~\eqref{fig:PionPole}.
As $p'$ varies along the rotated axis, these branch points trace out boundaries that the $p''$ contour cannot cross, as indicated by the solid red lines in the figure. 
This configuration does not pose a problem because the $p''$ contour runs parallel to these boundaries.

To regularize the ultraviolet behavior of the potential, we introduce the following separable regulator:
\begin{equation}
    V_{l'\Bar{l}'}(p', p) \to e^{-\frac{p'^4}{\Lambda^4}} V_{l'\Bar{l}'}(p', p) e^{-\frac{p^4}{\Lambda^4}}\, , 
    \label{eqn:DefReg}
\end{equation}
where $\Lambda$ denotes the ultraviolet momentum cutoff.
For large values of $p$ and $p'$, the regulator takes the following asymptotic form:
\begin{equation}
    \propto e^{-\cos{4\theta} \, \frac{p^4}{\Lambda^4}}\, e^{-\ii\sin{4\theta}\,\frac{p^4}{\Lambda^4}}\, .    
\end{equation}
The chosen value of $\theta$ is usually small enough that $\cos{4\theta} \,(p^4/\Lambda^4)$ remains positive, which in turn ensures the proper ultraviolet regularization of $t(p', p, E_2)$.

\begin{figure}[tb]
    \centering
    \includegraphics[scale=0.8]{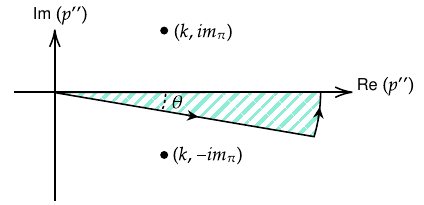}
    \caption{The analytic structure of $V^{1\pi}(k,p'')$ with the $p''$ contour rotated by an angle $\theta$.
The solid dots indicate the end-point singularities of the integral.}
    \label{fig:OnShell_PionPole}
\end{figure}

Although we do not need the on-shell $t$-matrix $t(k, k, E_2)$ in this paper, where $k = \sqrt{m_N E_2}$ is real, it can be calculated from the off-shell solution by using the known off-shell $t(p'', p, E_2)$ as input on the right-hand side of Eq.~\eqref{eqn:LSEOnJacobi}.
In this contour integration, the branch point $(k, -\ii m_\pi)$ does not cross the deformed contour provided that 
$k< m_{\pi}/\tan{\theta}$, as illustrated in Fig.~\ref{fig:OnShell_PionPole}. 
This condition imposes an upper limit on the accessible values of $k$, which is nevertheless sufficiently high for ChEFT applications, where the on-shell momenta under consideration are typically $\lesssim 3 m_\pi$.

The Faddeev equation in momentum space is given by:
\begin{align} 
&T(p' q' \alpha' ; \phi)\nonumber\\
& =\sum_{\Bar{\alpha}', l_d}   \int_{-1}^1 dx\; t_{l'\Bar{l'}} \left[p', \pi_1(q',q_0), E_3 - \frac{3q'^2}{4m_N}\right]  \dfrac{\varphi_{l_d}\Big [\pi_2(q',q_0)\Big ]G_{{\Bar{\alpha}'} \alpha_d} (q'\, q_0\,x)}{\pi_1^{\Bar{l}'}(q',q_0) \, \pi_2^{l_d}(q',q_0)}\nonumber\\
&\quad +\sum_{\Bar{\alpha}',\alpha''}  
\int_0^\infty \dd q'' {q''\,}^2\int_{-1}^1{\dd x}  \dfrac{G_{{\Bar{\alpha}}' \alpha''}(q'\, q''\,x)}{\pi^{\Bar{l}'}_1(q',q'') \, \pi_2^{l''}(q',q'')} \frac{t_{l'\Bar{l}'}\left[p', \pi_1(q',q''), E_3 -  \frac{3q'^2}{4m_N}\right]}{E_3  - \dfrac{q'^2 + {q''}^2 + x q'q''}{m_N}+\ii\epsilon}\nonumber\\
&\quad \times T\Big [\pi_2(q',q'') q'' \alpha''; \phi\Big ] \, .
    \label{eqn:compT}
\end{align}
Here, $\Bar{\alpha}$ denotes the channel coupled to $\alpha$ via the $NN$ interaction, obtained by replacing $l$ in $\alpha$ with $\Bar{l}$. $\phi$ is the collective symbol for the quantum numbers of the initial state, including $q_0$. 
The integral on the right-hand side of Eq.~\eqref{eqn:compT} exhibits three types of singularities that must be handled if all momenta are kept on the real axis, as detailed in Ref.~\cite{Glockle:1996jg}. 
First, the free propagator $G_0$ in the $q''$ integral has a pole when $E_3$ exceeds the $3N$ threshold.
Second, $t_{l'\Bar{l}'}(p', \pi_1, z)$ possesses a branch point at $z=0$, corresponding to the $q'$ satisfying $E_3 - 3q'^2/(4m_N) = 0$. This induces a singularity of $T(p'q'\alpha'; \phi)$ as a function of $q'$, which in turn produces a singularity of $T(\pi_2\, q'' \alpha''; \phi)$ in the $q''$ integral.
Third, in the ${}^3S_1-{}^3D_1$ channel, $t_{l'\Bar{l}'}(p', \pi_1, z)$ has an additional singularity: the deuteron pole at $z = E_d$.
If $E_3$ exceeds the nucleon-deuteron threshold, the deuteron pole corresponds to a singularity in $q'$ where $E_3 - 3q'^2/(4m_N) = E_d$, which again creates a singularity of $T(\pi_2\, q'' \alpha''; \phi)$ in the $q''$ integral.

In our implementation, these singularities are avoided by rotating $p'$, $q'$, and $q''$ in Eq.~\eqref{eqn:compT} into the complex plane, as defined by Eq.~\eqref{eqn:RotatedVar},
while the on-shell momentum $q_0$, the energies $E_d$ and $E_3$, and the Legendre variable $x$ remain real-valued.
The argument $\pi_2(q', q'')$ of $T(\pi_2 q'' \alpha''; \phi)$ also lies on the rotated contour. Consequently, only one spline operation is required to map $\pi_2(q', q'')$ onto the $p''$ mesh:
\begin{equation}
    T(\pi_2\, q''\, \alpha''; \phi) \approx \sum_m S_m(|\pi_2|) T(p''_m\, q''\, \alpha'';\phi)\, ,
\end{equation}
where $p''_m$ denote the rotated discrete momenta defined by Eq.~\eqref{eqn:RotMesh}, and the spline functions $S_m$ are given in Ref.~\cite{Glockle:1982agg}.
Using the momentum mesh and spline functions transforms the integral equation~\eqref{eqn:compT} into a linear system, which we denote symbolically by 
\begin{equation}
\sum_{mn\alpha''}K_{kr\alpha',mn\alpha''} \Psi_{mn\alpha''} = D_{kr\alpha'}
\, .\label{eqn:LinearEqn}
\end{equation}
Here, the indices $k$, $r$, $m$, and $n$ range over the momentum-mesh points associated with $p'$, $q'$, $p''$, and $q''$, respectively.
The kernel $K$, the unknown vector $\Psi$, and the driving term $D$ are defined as follows:
\begin{gather}
K_{kr\alpha',mn\alpha''} = w_m (p_m'')^2  w_n (q_n'')^2
\bra{p_k'q_n'\alpha'}1 - t P G_0\ket{p_m''q_n''\alpha''}\, ,\label{eqn:kernel}\\
\Psi_{mn\alpha''} = T(p_m''\, q_n''\, \alpha''; \phi) \, , \label{eqn:uknw_T}\\
D_{kr\alpha'}= \braket{p_k'\, q_r'\, \alpha' | t P | \phi} \, .\label{eqn:drvterm}
\end{gather}

The Faddeev breakup amplitude $T$ is used to compute the elastic scattering amplitude $U$. 
This is obtained through an additional integration:
\begin{align}
    &U_{\lambda ' I' , \lambda I}^{J}(q_0)\nonumber\\
    &=\sum_{l^{\prime}_d, l_d}  \int_{-1}^1 \dd x \Big (E_3-\frac{(x+2)q_0^2}{m_N} \Big )\dfrac{\varphi_{l_d^{\prime}}\left [\pi_1(q_0,q_0)\right ]}{\pi_1^{l_d'}(q_0,q_0)}\dfrac{ \varphi_{l_d}\left [\pi_2(q_0,q_0)\right ]}{\pi_2^{l_d}(q_0,q_0)} G_{\alpha_d'\alpha_d}(q_0q_0x)\nonumber\\
    &\quad + \sum_{l_d',\alpha''} \int q''^2\dd q'' \int_{-1}^1 \dd x \dfrac{\varphi_{l'_d}\left [\pi_1(q_0,q'')\right ]G_{\alpha_d' \alpha''}(q_0q''x)}{\pi_1^{l_d'}(q_0,q'')\;\pi_2^{l''}(q_0,q'')} T\left [\pi_2(q_0,q'')q''\alpha'';\phi_{\lambda I}^{J} \right ]\, .
    \label{eqn:NdScatAmp2NF}
\end{align}
$U_{\lambda ' I' , \lambda I}^{J}(q_0)$ is then transformed into $U_{\lambda ' \Sigma' , \lambda \Sigma}^{J}(q_0)$ via Eq.~\eqref{eqn:SigmaITrans}.
The integration contour for $q''$ in Eq.~\eqref{eqn:NdScatAmp2NF} is also rotated; consequently, the argument $\pi_2(q_0, q'')$ of $T$ in the second line does not lie on the rotated contour for $p'$.
To evaluate $T[\pi_2(q_0,q'')q''\alpha'';\phi_{\lambda I}^{J}]$, which is then fed into Eq.~\eqref{eqn:NdScatAmp2NF}, we again use Eq.~\eqref{eqn:compT} by setting $p' = \pi_2(q_0,q'')$.
Therefore, two additional integration steps are required to compute $U_{\lambda ' I' , \lambda I}^{J}(q_0)$ after obtaining the initial solution $T(p'_n q'_m \alpha'; \phi)$ via Eq.~\eqref{eqn:compT}, where $p'_n$ and $q'_m$ are the rotated meshes defined by Eq.~\eqref{eqn:RotMesh}.

\section{Perturbation theory for subleading order\label{sec:PertNLO}}

A key feature of our Faddeev-equation implementation is the perturbative treatment of subleading-order EFT interactions. We adopt the power counting of chiral nuclear forces as presented in Refs.~\cite{Long:2011xw, Long:2012ve,
Wu:2018lai}.
Two types of soft scales arise in EFTs. The first comprises nucleon momenta, e.g., the initial and final momenta of the nucleon and deuteron, and the deuteron binding momentum $\gamma_d \equiv \sqrt{m_N B_d} \simeq 46$ MeV.
The second consists of dimensionful parameters encoded in LECs, such as the $NN$ scattering lengths and effective ranges.
We use the pion mass $m_\pi$ as a generic proxy for these scales.
The breakdown scale is chosen to be the nucleon-delta mass splitting $\delta = 293$ MeV.

At LO, where the potential must be treated nonperturbatively --- that is, where the Faddeev equation, Eq.~\eqref{eqn:compT}, is solved exactly --- we include OPE potentials in the $\cs{1}{0}$, $\csd$, and $\chp{3}{0}$ partial waves, where contact potentials provide the necessary short-range interactions alongside the OPE.
In all other partial waves, OPE is sufficiently weak to be treated as an NLO correction.
For $\cs{1}{0}$ and $\csd$, the contact potential is expressed in momentum space as
\begin{equation}
 \braket{p'; lsj | V_S^{(0)} | p; lsj} = C_{0}^{(0)} \,
\end{equation}
where the superscript ``(0)'' indicates LO, ``(1)'' denotes NLO, and so on.
We suppress the channel label on the LECs when there is no risk of confusion.
At $\chp{3}{0}$, the contact interaction takes the following form:
\begin{equation}
 \braket{p';\chp{3}{0} | V_S^{(0)} | p;\chp{3}{0}} = C_0^{(0)} p' p \,.
\end{equation}

At NLO, OPE begins to contribute to additional channels, up to a maximum orbital angular momentum of $l \leqslant 2$.
We also retain the higher partial waves coupled to these channels; accordingly, $\cpf$ and $\cdg$ are included.
In addition, the $\cs{1}{0}$ potential receives the following NLO correction~\cite{Long:2012ve}:
\begin{equation}
 \braket{p';\cs{1}{0} | V_S^{(1)} | p;\cs{1}{0}}
 = C_0^{(1)} + \frac{D_0^{(1)}}{2} ({p'}^2 + p^2) \,,
\end{equation}
where $C_0^{(1)}$ is the NLO correction to the LO LEC $C_0^{(0)}$, and $D_0^{(1)}$ is the LEC associated with the momentum-dependent $\cs{1}{0}$ contact interaction.

In summary, our chiral potentials contain three undetermined LECs at LO and one at NLO.
These LECs are fixed by fits to the phase shifts from the Nijmegen partial-wave analysis~\cite{NNonline, Stoks:1993tb} up to $k=300$ MeV, with $k$ the center-of-mass momentum.
In addition, the deuteron binding energy, $B_d=2.225$ MeV, is reproduced at LO and NLO.

A crucial feature of this power counting for implementing the Faddeev equation is the significantly smaller number of channels at LO compared to subleading orders, as shown in Table~\ref{tab:NNChannels}.
We leverage this feature to implement a perturbative treatment of the subleading interactions.
To this end, the full space of $3N$ channels, $\mathscr{C}$, is decomposed into two sets.
The first, $\mathscr{A}$, consists of Jacobi partial waves whose two-body subsystem is subject to the LO interactions $\cs{1}{0}$, $\csd$, and $\chp{3}{0}$.
The remaining channels form the complementary set
$\mathscr{B}$:
\begin{equation}
    \mathscr{C} = \mathscr{A} \oplus \mathscr{B}\, .
\end{equation}
For any channel $\beta$ in $\mathscr{B}$ and any other 3N channel $\gamma$, we have
\begin{equation}
    \braket{p'q'\beta | V^{(0)} | p q\gamma} = 0\, .
    \label{eqn:betaV0Vanish}
\end{equation}
As we shall see shortly, this seemingly trivial identity proves useful for perturbative calculations.

\begin{table}[tb]
\centering
 \setlength\extrarowheight{5pt}
\setlength{\tabcolsep}{8pt}
\begin{tabular}{@{}llll@{}}
\toprule
\toprule
Order  & $NN$ partial waves\\ 
\midrule
\multirow{1}{*}{LO} & $\cs{1}{0}$, \,$\csd$, \,$\chp{3}{0}$ \\
\midrule
\multirow{1}{*}{NLO} &$\cs{1}{0}$,\, $\chp{1}{1}$,\, $\chp{3}{1}$,\, $\cpf$,\, $\cd{1}{2}$,\, $\cd{3}{2}$,\, $\cdg$\\
\bottomrule
\bottomrule
\end{tabular}
\caption{The $NN$ channels for the LO and NLO potentials.}
\label{tab:NNChannels}
\end{table}

To treat the NLO potential perturbatively, we begin with the formal EFT expansions of the two-body potential $V$, the two-body $t$-matrix $t$, the three-body Faddeev breakup operator $T$, and the initial state $\phi$:
\begin{align}
    V &= V^{(0)} + V^{(1)} + \cdots \, , \\
    t &= t^{(0)} + t^{(1)} + \cdots \, , \\
    T &= T^{(0)} + T^{(1)} + \cdots \, , \\
    \phi &= \phi^{(0)} + \phi^{(1)} + \cdots \, .
\end{align}
Substituting these expansions into Eqs.~\eqref{eqn:LSE} and \eqref{eqn:Faddeev} yields the following perturbative hierarchy:
\begin{gather}
T^{(0)}\NdKet{0} = t^{(0)}P\NdKet{0}+t^{(0)}PG_0T^{(0)}\NdKet{0}\, ,\label{eqn:T0phi0} \\
T^{(0)}\NdKet{1} = t^{(0)}P\NdKet{1}+t^{(0)}PG_0T^{(0)}\NdKet{1}\, ,\label{eqn:T0phi1} \\
T^{(1)}\NdKet{0} = t^{(1)}P\NdKet{0}+t^{(1)}PG_0T^{(0)}\NdKet{0}+t^{(0)}PG_0T^{(1)}\NdKet{0}\, ,
\label{eqn:T1phi0}
\end{gather}
where the $t^{(1)}$ is given by
\begin{equation}
t^{(1)}=V^{(1)}+V^{(1)}G_0t^{(0)}+V^{(0)}G_0t^{(1)}\, .
\end{equation}
This result should be compared with the direct calculation of $T^{(1)}$. 
\begin{equation}
T^{(1)}\ket{\phi^{(0)}} = 
(1+T^{(0)}G_0)t^{(1)}P(1+G_0T^{(0)}) \ket{\phi^{(0)}}
\, .
\end{equation}
Equations \eqref{eqn:T0phi0}, \eqref{eqn:T0phi1}, and \eqref{eqn:T1phi0} share the same integral kernel.
\begin{equation}
    K \equiv 1-t^{(0)}PG_0\, ,\label{eqn:kernel_FKP}
\end{equation}
but differ in their driving terms,
\begin{align}
D^{(0)} &\equiv t^{(0)}P\ket{\phi^{(0)}}\, ,\label{eqn:drvtm_lo}\\
D^{(0,1)} &\equiv t^{(0)}P\NdKet{1} \, ,\\
D^{(1,0)} &\equiv t^{(1)}P\left(1 + G_0T^{(0)}\right) \ket{\phi^{(0)}} \, .
\label{eqn:drvtm_nlo}
\end{align}

By Eq.~\eqref{eqn:betaV0Vanish}, the operator $t^{(0)}PG_0$, acting on the right, annihilates any $3N$-channel state $\beta \in \mathscr{B}$:
\begin{equation}
\braket{p'q' \beta | t^{(0)}PG_0 |pq\alpha}=0\, .
\end{equation}
As a result, the kernel has a block-triangular structure:
\begin{align} 
K = \left(
\begin{array}{c|c}
K_{A} & K_{AB} \\
\hline 
\mathbf{0} & \mathbf{1}
\end{array}
\right)\, ,
\end{align} 
where $K_A$ acts only within $\mathscr{A}$ and $K_{AB}$ couples $\mathscr{A}$ to $\mathscr{B}$.
\begin{align}
    &{K_A}(\alpha', \alpha) = \braket{\alpha'| K |\alpha} \quad \alpha, \alpha' \in \mathscr{A}\, ,\\
    &K_{AB}(\alpha', \beta) = \braket{\alpha'| K |\beta} \quad \beta \in \mathscr{B}\, .
\end{align}
Here, and later when it does not cause confusion, we omit the labels for the momentum mesh.
Upon discretization, $K_A$ is typically smaller than the full $K$, thus saving computational resources.
Accordingly, the unknown vector $\Psi$ and the driving term $D$ are also split into two parts:
\begin{align}
    \Psi^T &= (\Psi_A, \Psi_B) \, ,\\
    D^T &= (D_A, D_B)\, ,   
\end{align}
where the unknown vectors at each order are defined as follows:
\begin{align}
    \Psi^{(0)} &\equiv T^{(0)}\NdKet{0} \, , \\
    \Psi^{(0, 1)} &\equiv T^{(0)}\NdKet{1} \, , \\
    \Psi^{(1, 0)} &\equiv T^{(1)}\NdKet{0} \, .
\end{align}
With this decomposition, we can write Eqs. \eqref{eqn:T0phi0} and \eqref{eqn:T0phi1} as:
\begin{align}
\begin{cases}
    K_A \Psi_A^{(0)} &= D^{(0)}_A \, ,\\
    \Psi_B^{(0)} &= D^{(0)}_B = 0  \, . 
\end{cases}
\end{align}
and
\begin{align}
\begin{cases}
    K_A \Psi_A^{(0, 1)} &= D^{(0, 1)}_A\, , \\
    \Psi_B^{(0, 1)} &= D^{(0, 1)}_B = 0\, .
\end{cases}
\end{align}
The case corresponding to Eq.~\eqref{eqn:T1phi0} is more involved.
\begin{align}
\begin{cases}
K_A \Psi_A^{(1, 0)} = D^{(1, 0)}_A - K_{AB} D^{(1, 0)}_B\, , \\
    \Psi^{(1, 0)}_B = D^{(1, 0)}_B \, ,   
\end{cases}
\end{align}
where for $\alpha' \in \mathscr{A}$,
\begin{gather}
D^{(1, 0)}_A(\alpha') = \braket{\alpha' | t^{(1)}P\left(1 + G_0T^{(0)}\right) | \phi^{(0)}} 
\, ,\\
K_{AB} D^{(1, 0)}_B(\alpha') = - \sum_{\beta \in \mathscr{B}} \braket{\alpha' | t^{(0)} P G_0 | \beta} \braket{\beta |t^{(1)}P\left(1 + G_0T^{(0)}\right) |\phi^{(0)}}\, .
\end{gather}
Therefore, even at NLO we deal with a linear system that spans only $\mathscr{A}$. While the integral kernel remains identical to that at LO, the main computational effort at NLO shifts to constructing the driving term. By mathematical induction, one finds that at all higher orders the same kernel is reused, although the driving terms become increasingly complex. Because of this feature, we refer to this approach to perturbative calculations as fixed-kernel perturbation theory (FKPT).

In all implementations in this work, $\mathscr{A}$ comprises 6 channels for $J=\frac{1}{2}$ and 8 for $J\geqslant\frac{3}{2}$. The dimension of the full space of $3N$ channels $\mathscr{C}$ varies with $J$ as follows: 22 for $J=\frac{1}{2}$, 38 for $J=\frac{3}{2}$, 46 for $J=\frac{5}{2}$, $\cdots$. In general, the dimension of the LO kernel is reduced by a factor of $\left( \frac{\text{Dim}_{\mathscr{A}}}{\text{Dim}_{\mathscr{C}}}\right)^2$ compared to that of the full kernel; consequently, the former is at least an order of magnitude smaller than the latter.

Using $T^{(0)}\ket{\phi^{(0)}}$, $T^{(0)}\ket{\phi^{(1)}}$, and $T^{(1)}\ket{\phi^{(0)}}$ as inputs, we can express the elastic scattering amplitude at LO and NLO as
\begin{gather}
U^{(0)}=\braket{\phi'^{(0)} | PG_0^{-1} + PT^{(0)}|\phi^{(0)}} 
\, ,\label{eqn:U_LO} \\
U^{(1)}=\bra{\phi'^{(1)}}PG_0^{-1}+PT^{(0)}\ket{\phi^{(0)}}+\bra{\phi'^{(0)}}PT^{(1)}\ket{\phi^{(0)}}\nonumber\\
 +\bra{\phi'^{(0)}}PG_0^{-1}+PT^{(0)}\ket{\phi^{(1)}}\, \label{eqn:U_NLO}\, .
\end{gather}
In the chiral power counting adopted in this work, the $\csd$ interaction vanishes at NLO, as noted in Table~\ref{tab:NNChannels}.
As a result, the NLO correction to the deuteron wave function is also zero, which implies that the NLO corrections to the initial and final $Nd$ states vanish as well:
\begin{equation}
    \ket{\phi^{(1)}}=0\, .
\end{equation}

Although exact unitarity is violated in perturbation theory, we can still extract the NLO phase shifts from $U^{(1)}$ in a manner consistent with power counting.
As detailed for $NN$ coupled channels (see, e.g., Ref.~\cite{Long:2011xw}), the phase-shift parameters are expanded as
\begin{align}
    \delta_n&=\delta_n^{(0)}+\delta_n^{(1)}+\cdots \, ,
\end{align}
where $n$ indexes the various phase-shift parameters, including mixing angles.
The exact relation between $U$ and $\delta_n$ is provided in Ref.~\cite{Seyler:1969sii} and can be written symbolically as follows:
\begin{equation}
    U=W(\delta_1, \delta_2, \cdots)\, .
\end{equation}
We expand both sides:
\begin{align}
    U^{(0)} + U^{(1)} + \cdots = W(\delta_1^{(0)}, \delta_2^{(0)}, \cdots) + \sum_n \frac{\partial W}{\partial \delta_n} \delta^{(1)}_n + \cdots\, .
\end{align}
Using $U^{(1)}$ from our perturbative calculations as input, together with the matrix $\partial W/\partial \delta_n$, we can extract $\delta_n^{(1)}$ by matching orders on both sides.

\section{Benchmarks\label{sec:Bench}}

To determine the number of one-dimensional momentum mesh points, $N$, and the contour-rotation angle, $\theta$, we compare several options.
Table~\ref{tab:CnvgeWithNP} and Table~\ref{tab:CnvgeWithTheta} show the effects of varying $N$ and $\theta$, respectively, on the $^2S_{\frac{1}{2}}$ phase shifts.
The variations are generally at the subpercent level or smaller, which is more than sufficient for our purposes.
For definiteness, we adopt $N=48$ mesh points and a contour-rotation angle of $10^\circ$.

\begin{table}[ht]
\centering
\setlength\extrarowheight{5pt}
\setlength{\tabcolsep}{8pt}
\begin{tabular}{@{}lcccccc@{}}
\toprule
\toprule
\multirow{2}{*}{\quad $E_N$} & \multicolumn{2}{c}{3 MeV} & \multicolumn{2}{c}{14 MeV} & \multicolumn{2}{c}{30 MeV} \\
\cmidrule(lr){2-3} \cmidrule(lr){4-5} \cmidrule(lr){6-7}
& Re & Im & Re & Im & Re & Im \\
\midrule
\multirow{2}{*}{$N=32$} 
& -12.61 & -0.04 & -41.48 & 21.69 & -86.47 & 41.14 \\
& -30.57 & 0.66 & -69.36 & 25.80 & -127.18 & 34.03 \\
\midrule
\multirow{2}{*}{$N=48$} 
& -12.64 & 0.00 & -41.42 & 21.78 & -86.55 & 41.02 \\
& -30.95 & 0.00 & -69.35 & 25.99 & -127.17 & 33.81 \\
\midrule
\multirow{2}{*}{$N=64$} 
& -12.64 & 0.00 & -41.42 & 21.78 & -86.55 & 41.02 \\
& -30.97 & 0.00 & -69.35 & 26.00 & -127.16 & 33.82 \\
\bottomrule
\bottomrule
\end{tabular}
\caption{Real (Re) and imaginary (Im) parts of the $^2S_{\frac{1}{2}}$ phase shifts (in degrees) are shown for various incident nucleon energies $E_N$. The number of mesh points, $N$, is varied, with a fixed contour-rotation angle $\theta = 10^{\circ}$. For each $N$, the first and second rows correspond to the LO and NLO phase shifts, respectively.}
\label{tab:CnvgeWithNP}
\end{table}

\begin{table}[ht]
\centering
\setlength\extrarowheight{5pt}
\setlength{\tabcolsep}{8pt}
\begin{tabular}{@{}lllllll@{}}
\toprule
\toprule
\multirow{2}{*}{\quad $E_N$} & \multicolumn{2}{c}{3 MeV} & \multicolumn{2}{c}{14 MeV} & \multicolumn{2}{c}{30 MeV} \\
\cmidrule(lr){2-3} \cmidrule(lr){4-5} \cmidrule(lr){6-7}
& Re & Im & Re & Im & Re & Im \\
\midrule
\multirow{2}{*}{$\theta=8^{\circ}$} 
& -12.62 & 0.00 & -41.42 & 21.76 & -86.58 & 41.04 \\
& -30.89 & 0.00 & -69.33 & 25.94 & -127.21 & 33.80 \\
\midrule
\multirow{2}{*}{$\theta=10^{\circ}$} 
& -12.64 & 0.00 & -41.42 & 21.78 & -86.55 & 41.02 \\
& -30.95 & 0.00 & -69.35 & 25.99 & -127.17 & 33.81 \\
\midrule
\multirow{2}{*}{$\theta=13^{\circ}$} 
& -12.64 & 0.00 & -41.43 & 21.79 & -86.56 & 41.02 \\
& -30.97 & 0.00 & -69.35 & 25.98 & -127.15 & 33.79 \\
\bottomrule
\bottomrule
\end{tabular}
\caption{
The $^2S_{\frac{1}{2}}$ phase shifts (in degrees) are shown for various values of the contour-rotation angle $\theta$, with the number of momentum mesh points fixed at $N = 48$. For each $\theta$, the first and second rows correspond to the LO and NLO phase shifts, respectively.}
\label{tab:CnvgeWithTheta}
\end{table}

To benchmark our contour-deformation implementation of the Faddeev equation, we compare our results with those obtained using the wave-packet continuum-discretization (WPCD) method~\cite{POMERANTSEV2016121}. We also note other studies in which WPCD has been successfully combined with various interactions to compute $Nd$ scattering~\cite{Miller:2021vby, Miller:2022beg, Zhai:2025grs}. For the interaction, we use LO potentials with a cutoff $\Lambda=400$ MeV.

We calculate phase shifts and mixing angles at nucleon laboratory energies $E_N = 3$, $14$, and $30$ MeV for $J^P = \frac{1}{2}^\pm$ and $\frac{3}{2}^+$. The results from the two methods are compared in Table~\ref{tab:BenchDelta_J0.5} for $J^P=\frac{1}{2}^\pm$ and in Table~\ref{tab:BenchDelta_J1.5+} for $J^P=\frac{3}{2}^+$. Overall, the discrepancies between the results of the two methods are less than one percent. This level of agreement is sufficient for the present study, as the subsequent NLO corrections and cutoff variations are expected to be substantially larger than the differences introduced by the numerical methods.

\begin{table}
\centering
\setlength\extrarowheight{5pt}
\setlength{\tabcolsep}{8pt}
\begin{tabular}{@{}lcccccc@{}}
\toprule
\toprule
\multirow{2}{*}{\quad $E_N$} & \multicolumn{2}{c}{3 MeV} & \multicolumn{2}{c}{14 MeV} & \multicolumn{2}{c}{30 MeV} \\
\cmidrule(lr){2-3} \cmidrule(lr){4-5} \cmidrule(lr){6-7}
& Re & Im & Re & Im & Re & Im \\
\midrule
\multirow{2}{*}{$^2D_{\frac{1}{2}}$} 
& -3.28 & 0.00 & -6.50 & 0.62 & -6.45 & 2.46 \\
& -3.27 & 0.00 & -6.49 & 0.63 & -6.29 & 2.46 \\
\midrule
\multirow{2}{*}{$^2S_{\frac{1}{2}}$} 
& -12.63 & 0.00 & -41.42 & 21.78 & -86.46 & 41.08 \\
& -12.88 & 0.00 & -41.59 & 21.89 & -86.95 & 40.56 \\
\midrule
\multirow{2}{*}{$\eta_{\frac{1}{2}}$} 
& 7.73 & 0.00 & 4.98 & -0.06 & 5.32 & -0.84 \\
& 7.48 & 0.00 & 4.95 & -0.06 & 5.30 & -0.83 \\
\midrule
\multirow{2}{*}{$^2P_{\frac{1}{2}}$} 
& -5.72 & 0.00 & 6.96 & 8.49 & 50.04 & 10.19 \\
& -5.70 & 0.00 & 7.02 & 8.51 & 50.29 & 10.27 \\
\midrule
\multirow{2}{*}{$^4P_{\frac{1}{2}}$} 
& 21.85 & 0.00 & 42.32 & 5.52 & 11.09 & 8.20 \\
& 21.81 & 0.00 & 42.39 & 5.56 & 11.19 & 8.22 \\
\midrule
\multirow{2}{*}{$\epsilon_{\frac{1}{2}}$} 
& 6.98 & 0.00 & 31.33 & 8.15 & -44.96 & 6.88 \\
& 6.99 & 0.00 & 31.45 & 8.19 & -44.73 & 6.98 \\
\bottomrule
\bottomrule
\end{tabular}
\caption{Real (Re) and imaginary (Im) parts of the phase shifts and mixing angles (in degrees) for $J^P=\frac{1}{2}^{\pm}$ at various incident nucleon energies.
For each phase shift, the upper rows show results obtained with the contour-deformation method, and the lower rows show results calculated with the WPCD method.}
\label{tab:BenchDelta_J0.5}
\end{table}

\begin{table}
\centering
\setlength\extrarowheight{5pt}
\setlength{\tabcolsep}{8pt}
\begin{tabular}{@{}lcccccc@{}}
\toprule
\toprule
\multirow{2}{*}{\quad $E_N$} & \multicolumn{2}{c}{3 MeV} & \multicolumn{2}{c}{14 MeV} & \multicolumn{2}{c}{30 MeV} \\
\cmidrule(lr){2-3} \cmidrule(lr){4-5} \cmidrule(lr){6-7}
& Re & Im & Re & Im & Re & Im \\
\midrule
\multirow{2}{*}{$^4S_{\frac{3}{2}}$} 
& -66.55 & 0.00 & -100.58 & 0.78 & -118.57 & 3.33 \\
& -66.56 & 0.00 & -100.34 & 0.66 & -118.03 & 3.16 \\
\midrule
\multirow{2}{*}{$^2D_{\frac{3}{2}}$} 
& 2.14 & 0.00 & 5.87 & 1.35 & 7.45 & 3.04 \\
& 2.13 & 0.00 & 5.88 & 1.36 & 7.48 & 3.07 \\
\midrule
\multirow{2}{*}{$^4D_{\frac{3}{2}}$} 
& -3.78 & 0.00 & -7.10 & 0.60 & -5.72 & 2.09 \\
& -3.73 & 0.00 & -7.09 & 0.61 & -5.72 & 2.11 \\
\midrule
\multirow{2}{*}{$\epsilon_{\frac{3}{2}}$} 
& 0.90 & 0.00 & 0.65 & 0.00 & -1.25 & 0.08 \\
& 0.90 & 0.00 & 0.65 & 0.00 & -1.27 & 0.08 \\
\midrule
\multirow{2}{*}{$\xi_{\frac{3}{2}}$} 
& 1.48 & 0.00 & 4.83 & -0.08 & 8.40 & -0.48 \\
& 1.48 & 0.00 & 4.83 & -0.09 & 8.42 & -0.48 \\
\midrule
\multirow{2}{*}{$\eta_{\frac{3}{2}}$} 
& -0.34 & 0.00 & -1.40 & -0.36 & -2.17 & -0.83 \\
& -0.34 & 0.00 & -1.40 & -0.36 & -2.18 & -0.82 \\
\bottomrule
\bottomrule
\end{tabular}
\caption{
Real (Re) and imaginary (Im) parts of the phase shifts and mixing angles (in degrees) for $J^P=\frac{3}{2}^{+}$ at various incident nucleon energies.
For each phase shift, the upper rows display values obtained by the contour-deformation method, while the lower rows display those computed with the WPCD method.
}
\label{tab:BenchDelta_J1.5+}
\end{table}

\begin{figure}
\centering
     \includegraphics[scale=0.26]{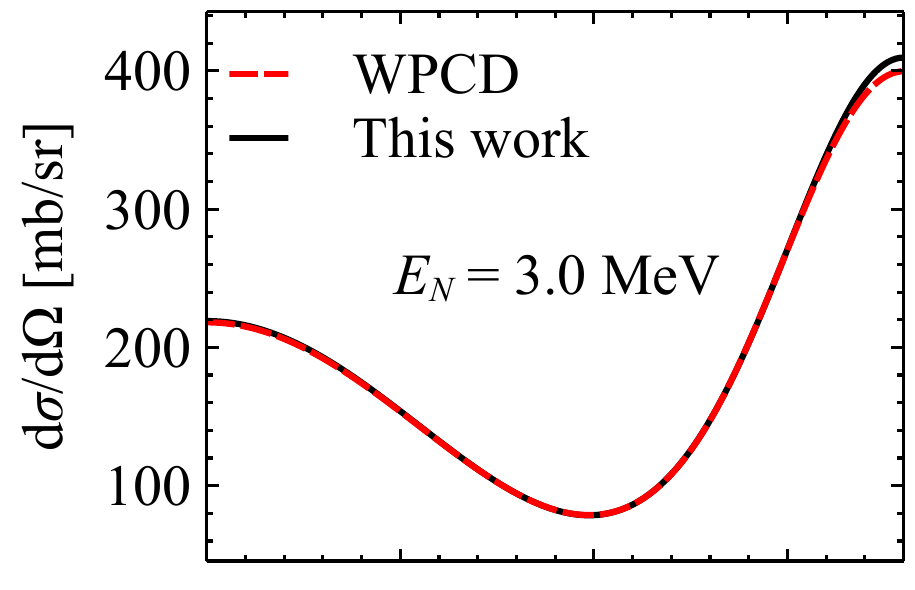}
     \includegraphics[scale=0.26]{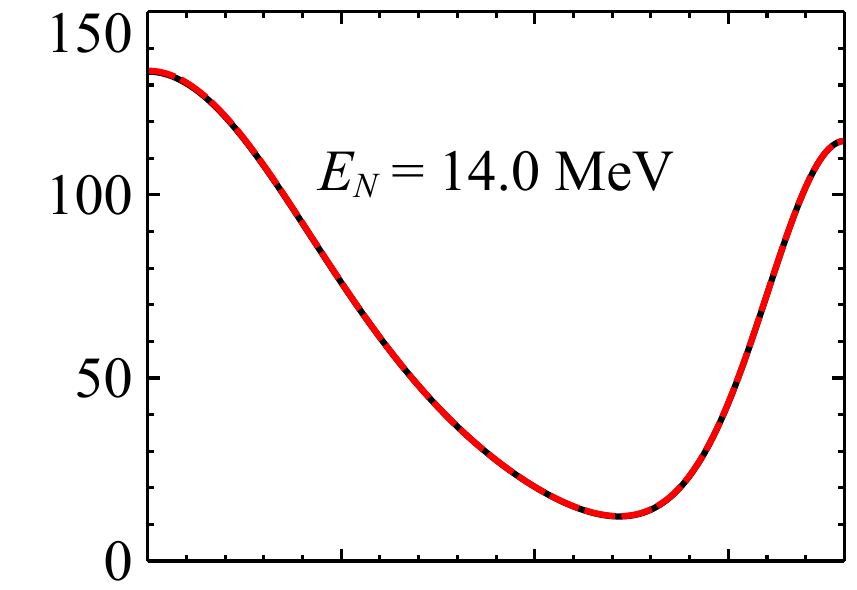}
     \includegraphics[scale=0.26]{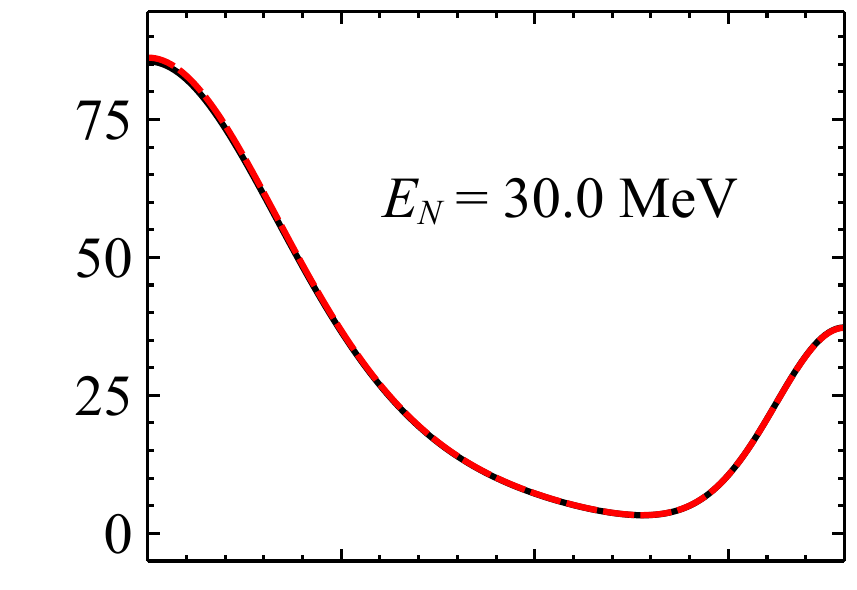}
     \includegraphics[scale=0.26]{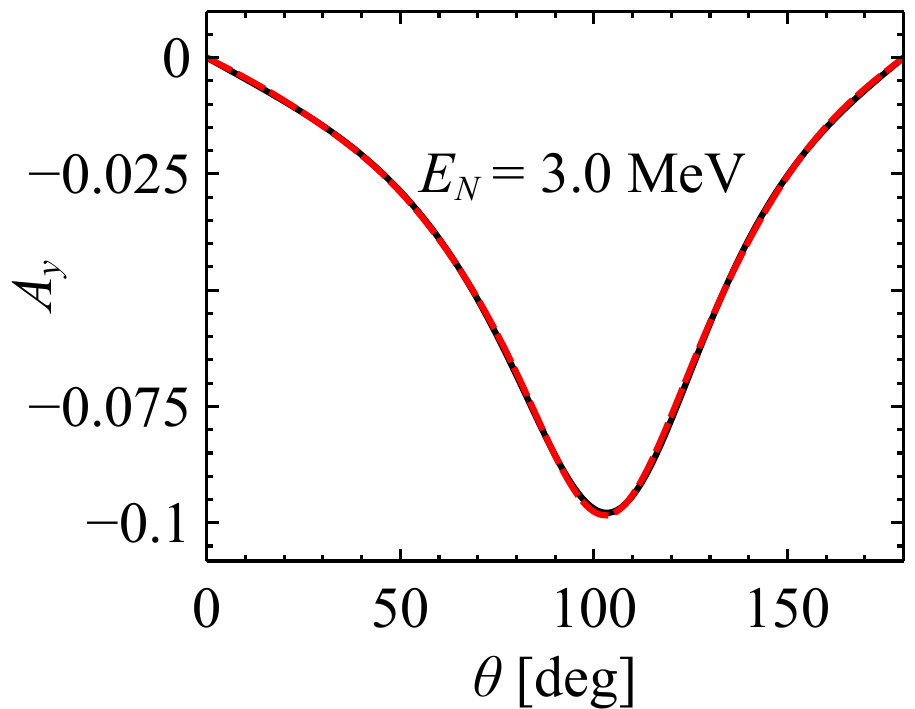}
     \includegraphics[scale=0.26]{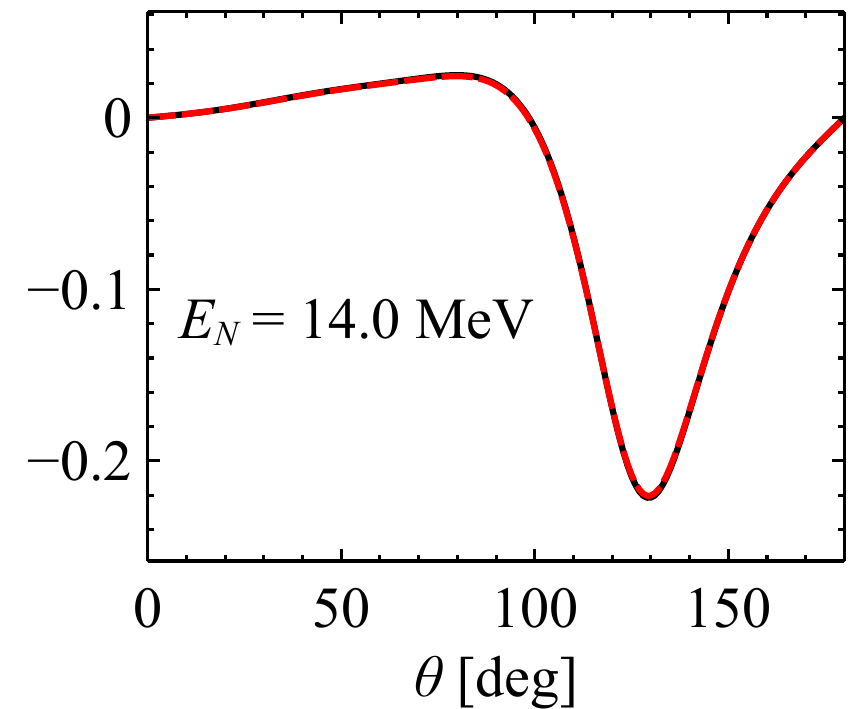}
     \includegraphics[scale=0.26]{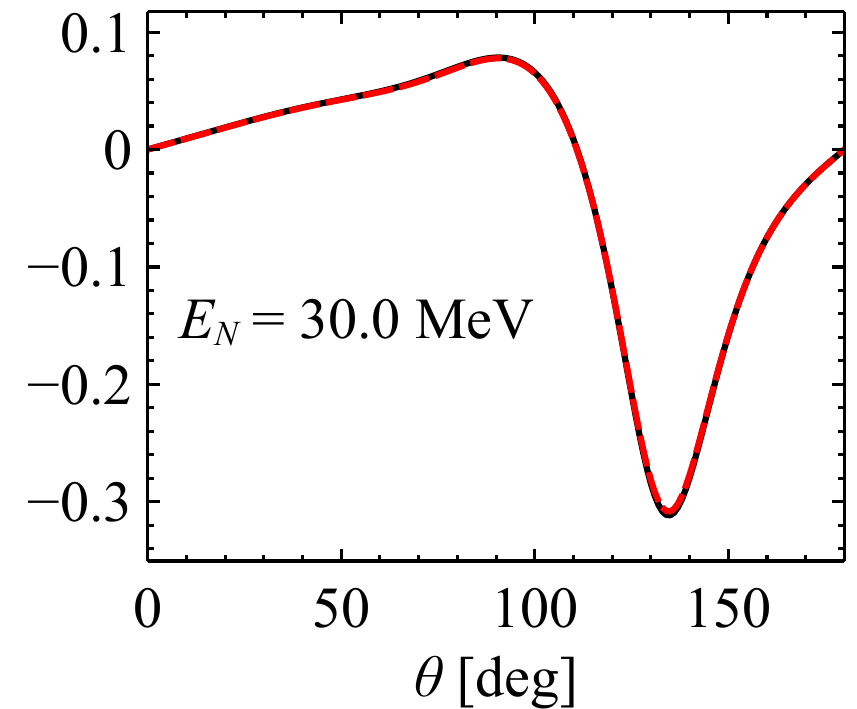}
     \caption{Comparison of the $Nd$ elastic-scattering differential cross sections and nucleon analyzing powers. Results obtained with the contour-deformation and WPCD methods are shown as solid and dashed curves, respectively.
     }\label{fig:BenchmarkWithWPCD}
\end{figure}

We also compare the differential cross sections and nucleon analyzing powers, $A_y$, calculated using the two methods, as shown in Fig.~\ref{fig:BenchmarkWithWPCD}.
Following Ref.~\cite{Glockle:1996jg}, we relate the partial-wave amplitude $U^J$ to these observables.
For these observables, the partial-wave sum of the scattering amplitude $U^J$ includes contributions up to $J^{P} \leqslant \frac{15}{2}^{\pm}$.
The level of agreement for these observables is consistent with that found for the phase shifts in Tables~\ref{tab:BenchDelta_J0.5} and \ref{tab:BenchDelta_J1.5+}.

Implementing the FKPT method — our perturbative treatment of NLO potentials outlined in Sec.~\ref{sec:PertNLO} — requires substantial extensions to the code beyond the standard Faddeev-equation solver.
To validate this functionality, we note that perturbation theory can be implemented straightforwardly within the nonperturbative framework, albeit with additional computational cost.
Starting from an auxiliary potential,
\begin{equation}
    V(x) = V^{(0)} + xV^{(1)}\, ,
    \label{eqn:AuxV2}
\end{equation}
where $x$ is an auxiliary multiplier sampled over an interval around 0. We can use the nonperturbative solver, i.e., Eqs.~\eqref{eqn:compT} and \eqref{eqn:NdScatAmp2NF}, to calculate the $x$-dependent scattering amplitude $U(E_3, x)$ for arbitrary $x$.
By numerically expanding $U(E_3, x)$ about $x = 0$, we obtain the NLO scattering amplitude in perturbation theory:
\begin{align}
    U(E_3, x)=U^{(0)}(E_3) + x U^{(1)}(E_3) +  \cdots \, .
\end{align}
This approach is computationally demanding, as it requires multiple evaluations of the amplitude at different values of $x$ to carry out the Taylor expansion numerically.
In practice, we use at least five $x$ values to ensure the numerical stability of the NLO results.
Compared with the FKPT method, the auxiliary-potential method requires at least an order of magnitude more computation time when both methods use the same number of mesh points $N$ and the same set of truncated $3N$ channels $\ket{\alpha}$.
Furthermore, the auxiliary-potential method has a large memory footprint because it does not exploit the significantly smaller size of the LO kernel.
While useful for benchmarking, these limitations motivate the adoption of the FKPT approach.
For the sampled phase shifts, the two approaches yield consistent results, agreeing to at least six significant digits.

\section{Results\label{sec:Results}}

We use $Nd$ elastic scattering to investigate two aspects of the power counting of chiral nuclear forces adopted in this paper. The LO and NLO potentials are explained in Sec.~\ref{sec:PertNLO}. One is the ultraviolet-cutoff dependence of the phase shifts, and the other is the effects of the NLO correction.

The ultraviolet cutoff $\Lambda$ introduced in Eq.~\eqref{eqn:DefReg} is arbitrarily chosen, and observables must be independent of its value up to the EFT uncertainty allowed at the given order; this is a manifestation of RG invariance. A violation of RG invariance can be interpreted as a defect in the power counting, which can be remedied by promoting certain operators to lower orders. In Ref.~\cite{Song:2016ale}, RG invariance of the triton binding energy and $Nd$ scattering lengths was observed, leading to the conclusion that the $3N$ forces are not required up to NLO for renormalization purposes.

\begin{figure}
    \centering
    \includegraphics[scale=0.3]{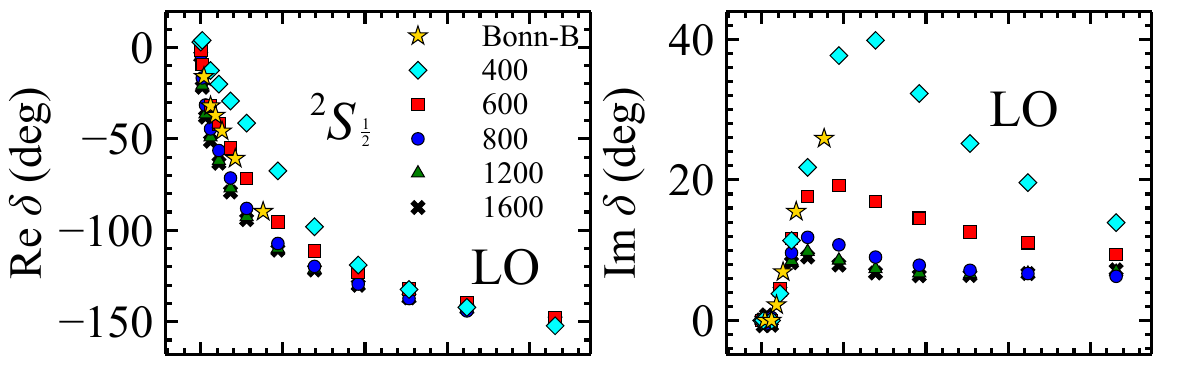}
    \includegraphics[scale=0.3]{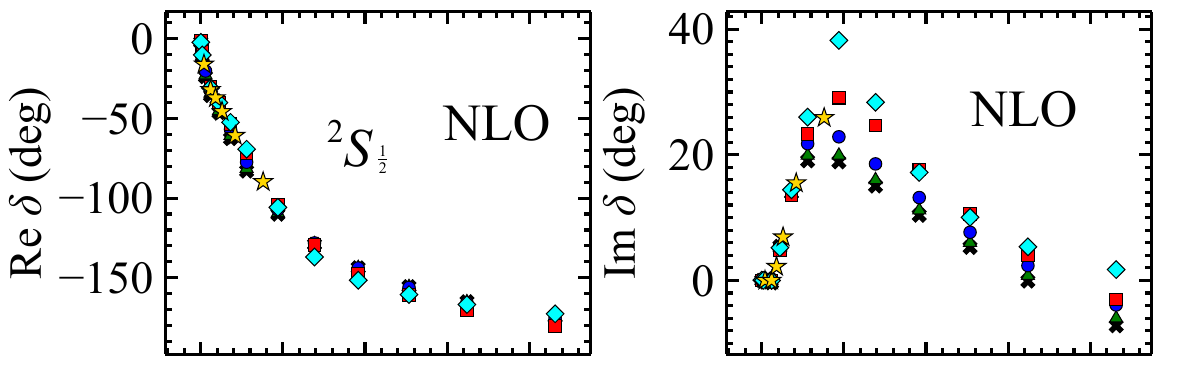}
    \includegraphics[scale=0.3]{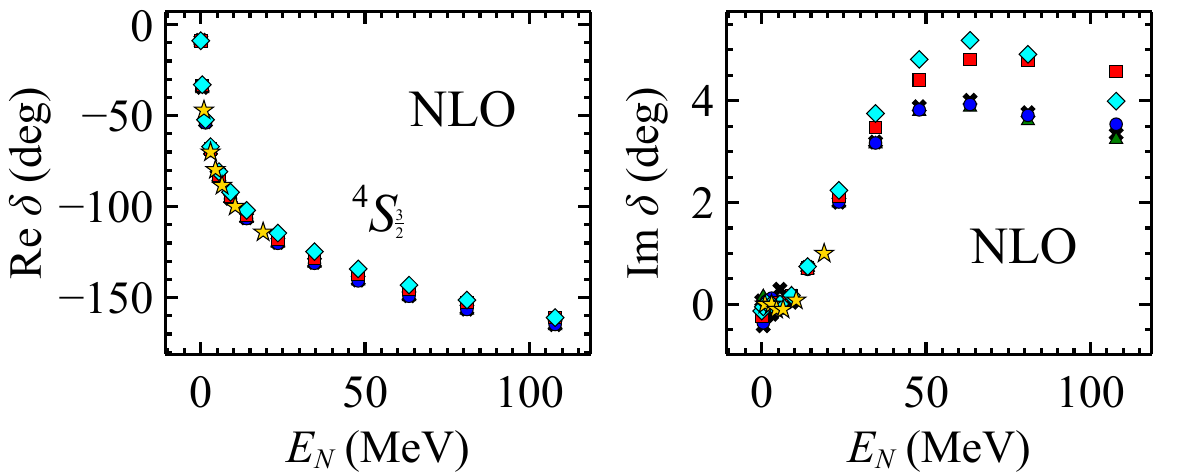}
    \includegraphics[scale=0.3]{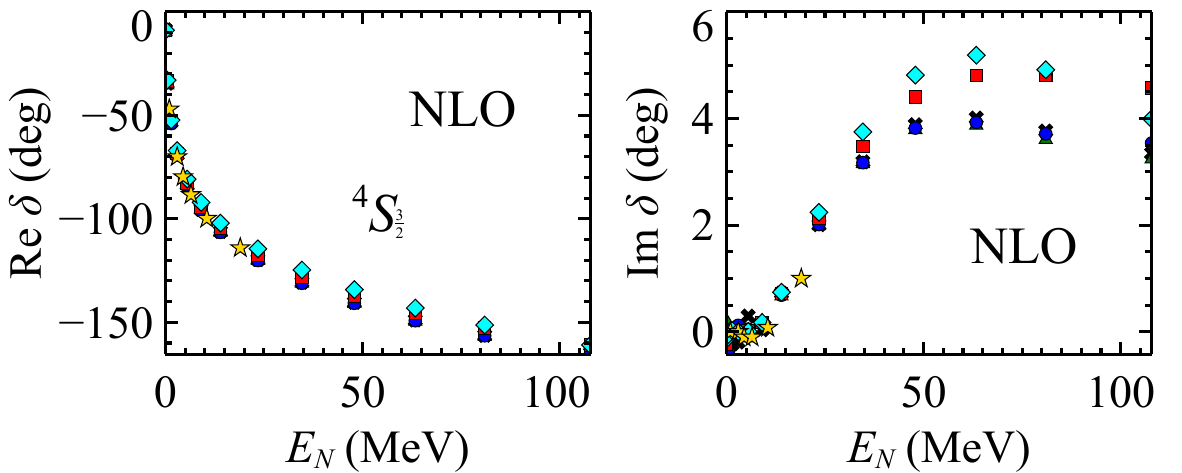}
    \caption{The doublet- and quartet-channel $S$-wave phase shifts as functions of $E_N$ for various values of the cutoff $\Lambda$. The top and bottom panels correspond to the $^2S_{\frac{1}{2}}$ and $^4S_{\frac{3}{2}}$ channels, respectively. The ``Bonn-B'' results are taken from Ref.~\cite{Born-B}.
    }
    \label{fig:2SAND4S_MMWLY}
\end{figure}

We verify this conclusion by inspecting how the phase shifts vary with $\Lambda$. Because ultraviolet divergences tend to be suppressed at higher orbital angular momentum, it suffices to study the $S$-wave phase shifts of $Nd$ scattering. When $\Lambda$ exceeds $\sim 600$ MeV in $\chp{3}{0}$ and $\sim 1000$ MeV in $\csd$, spurious $NN$ bound states begin to develop~\cite{Nogga:2005hy}. In order to avoid unphysical breakup thresholds in the $Nd$ system, we remove these spurious states using a method similar to that described in Ref.~\cite{Nogga:2005hy}. More concretely, we raise the energies of the spurious $NN$ states to a large positive value, for example a few GeV, so that they lie far outside the EFT region and do not open undesired reaction channels. This is implemented by adding the following to the LO two-body potential:
\begin{align}
    V^{(0)} + \Omega \ket{\Psi_s}\bra{\Psi_s} \, , 
\end{align}
where $\ket{\Psi_s}$ denotes the normalized spurious state, and $\Omega$ denotes the artificial positive energy assigned to it.
Figure~\ref{fig:2SAND4S_MMWLY} presents the phase shifts of $^2S_{\frac{1}{2}}$ and $^4S_{\frac{3}{2}}$ as $\Lambda$ varies from 400 to 1600 MeV.
Up to $E_N = 108$ MeV, the phase shifts exhibit convergence with respect to $\Lambda$.

\begin{figure*}
    \centering
        \centering
        \includegraphics[scale=0.28]{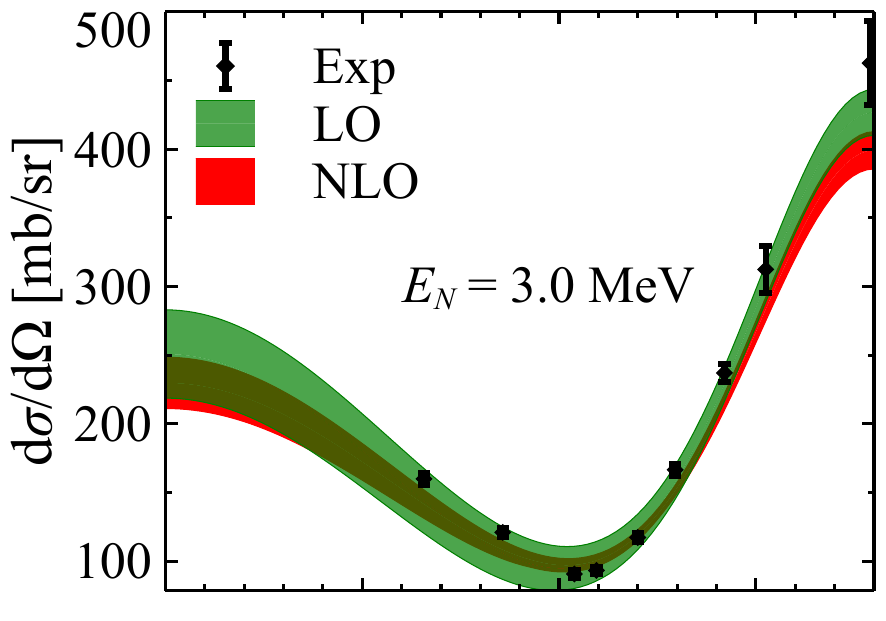}
        \includegraphics[scale=0.28]{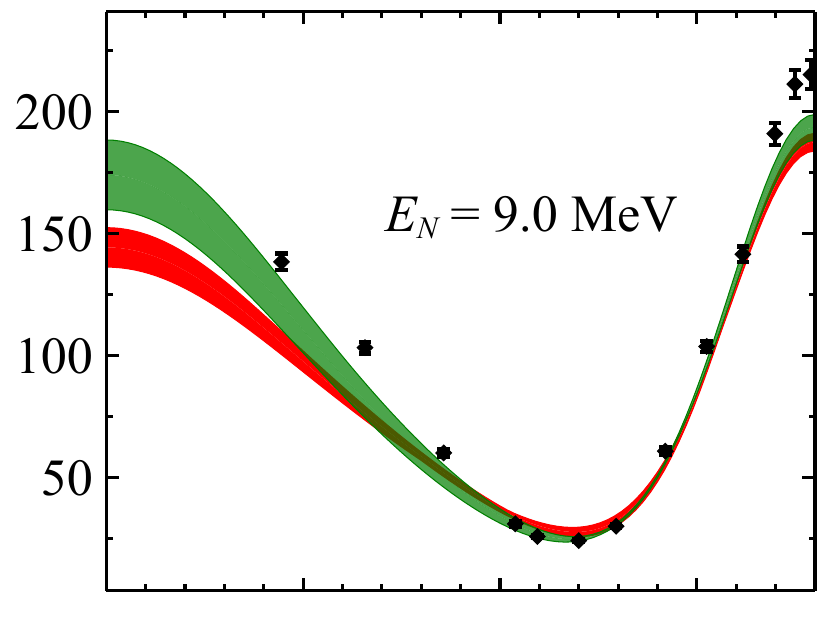}
        \includegraphics[scale=0.28]{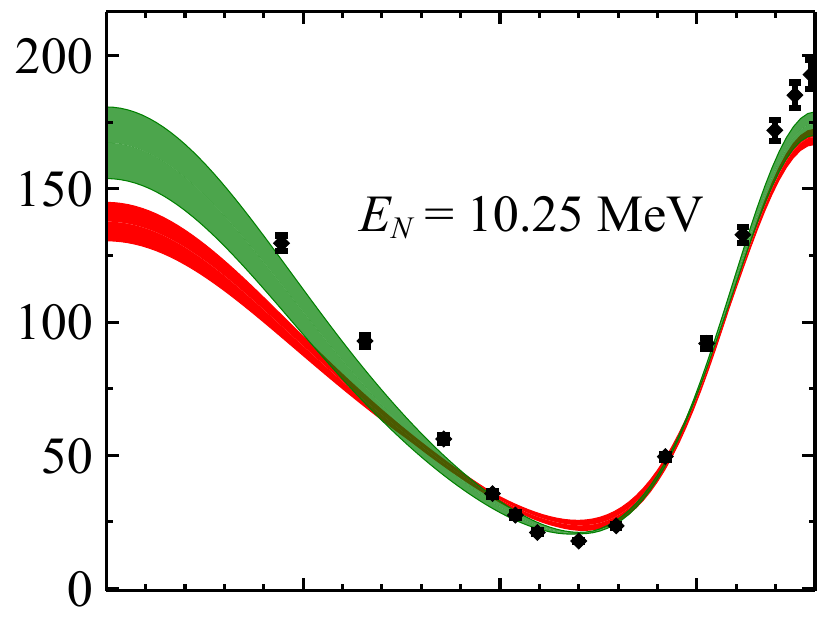}
        \includegraphics[scale=0.28]{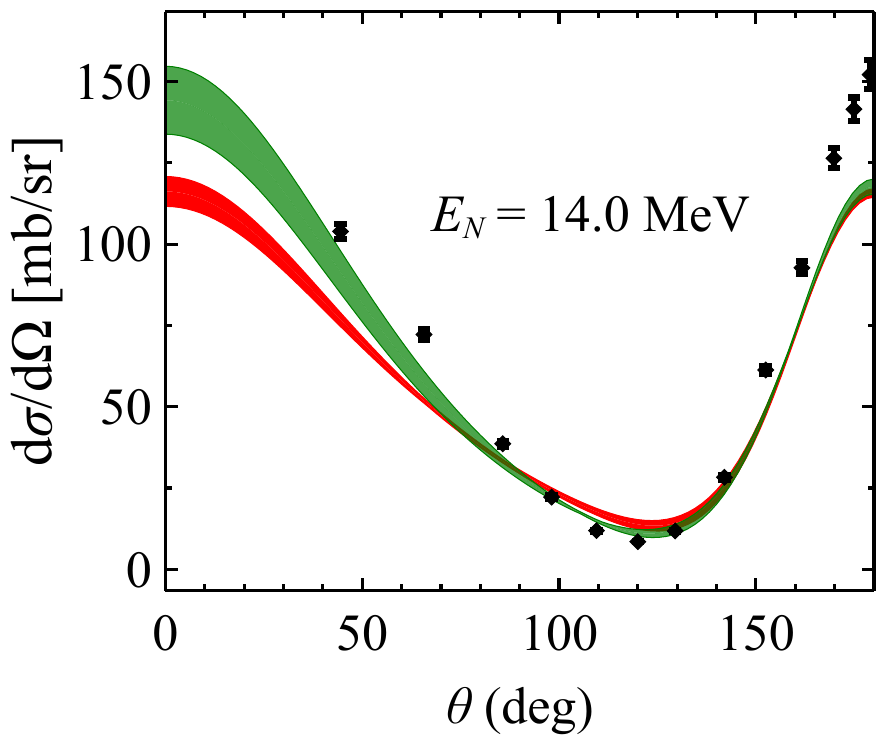}
        \includegraphics[scale=0.28]{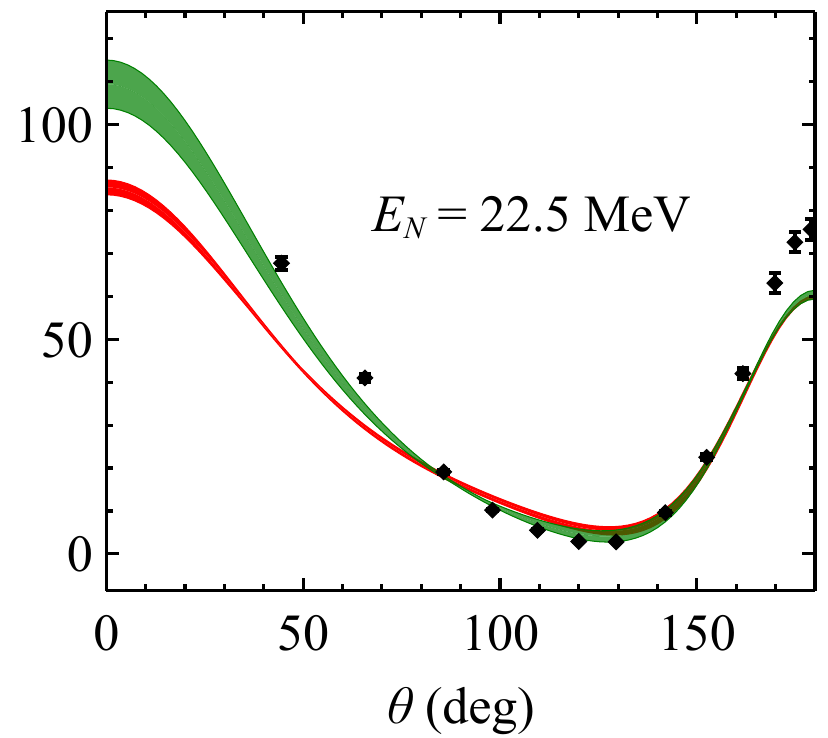}
        \includegraphics[scale=0.28]{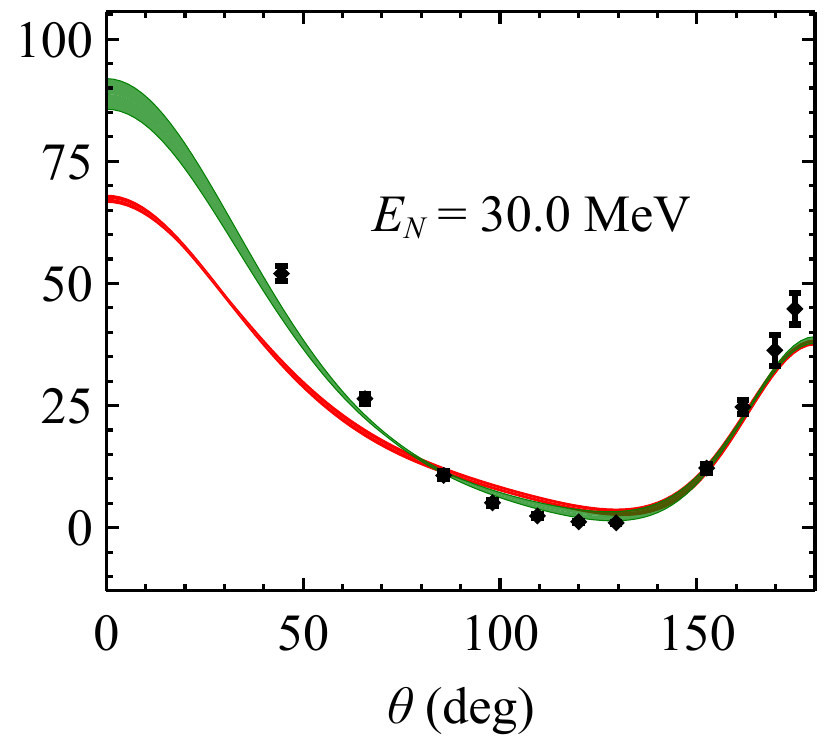}
        \caption{The LO and NLO angular distributions at various $E_N$ are shown. The bands are generated by varying $\Lambda$ from 400 to 1600 MeV. The $nd$ scattering data are taken from Ref.~\cite{Schwarz:1983rlz}.
        } 
        \label{fig:MMWDSG}
    \vspace{0.5cm}
\end{figure*}

\begin{figure*}
        \centering
        \includegraphics[scale=0.28]{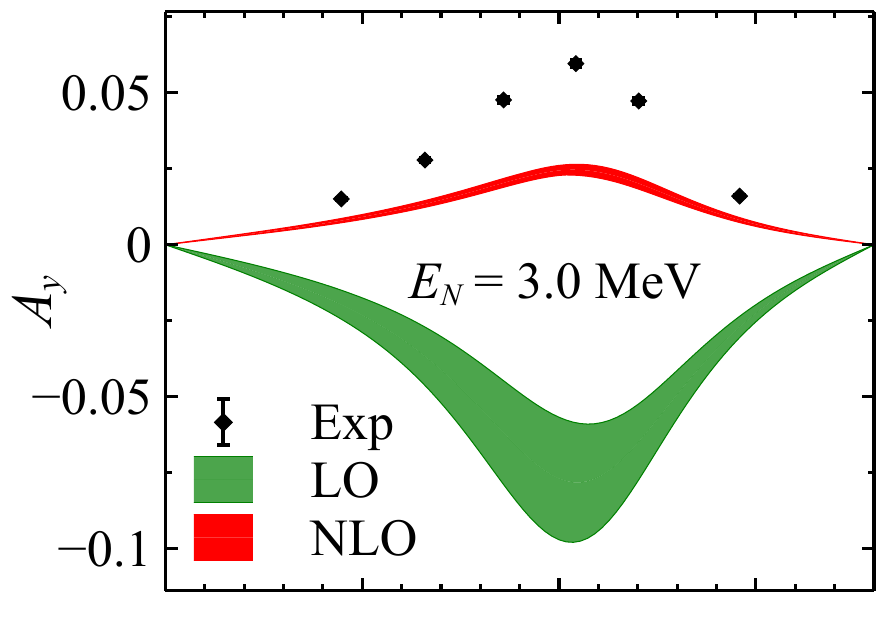}
        \includegraphics[scale=0.28]{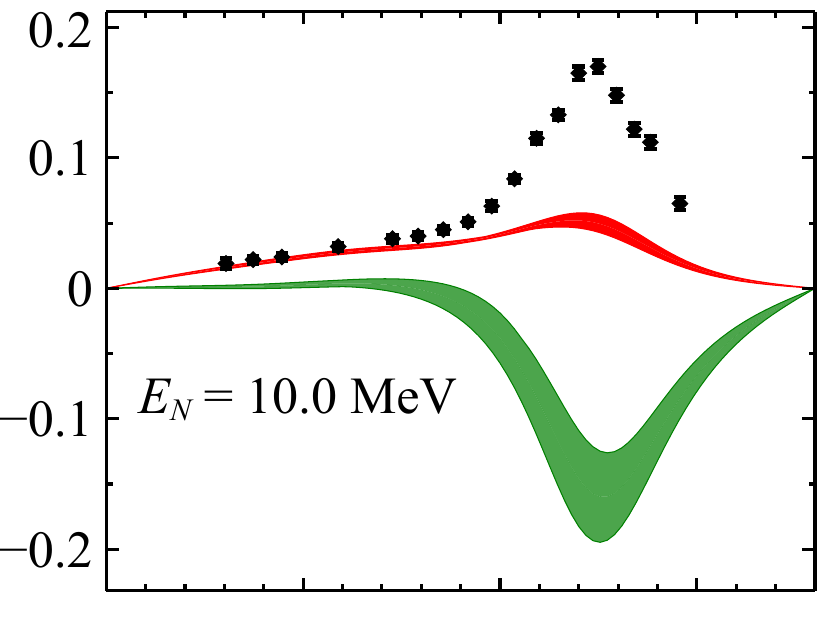}
        \includegraphics[scale=0.28]{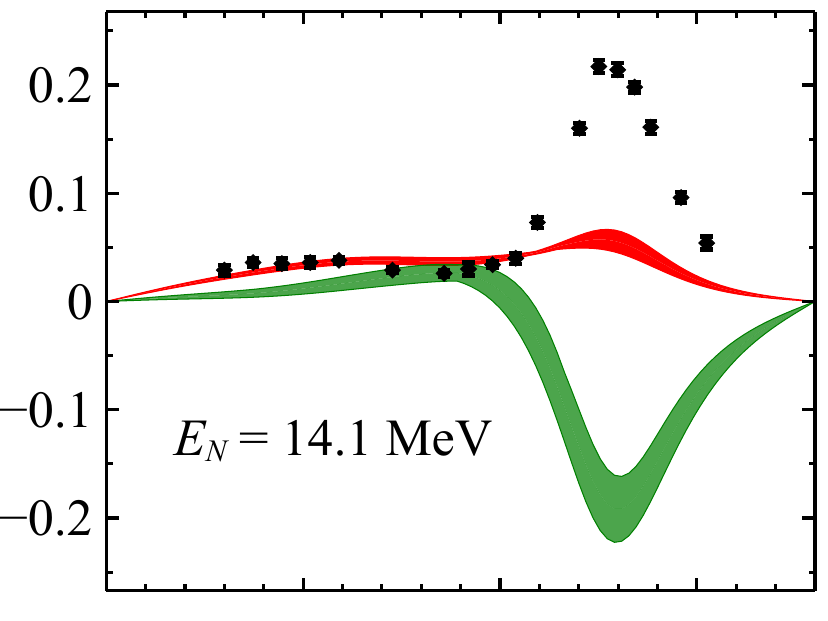}\\
        \includegraphics[scale=0.28]{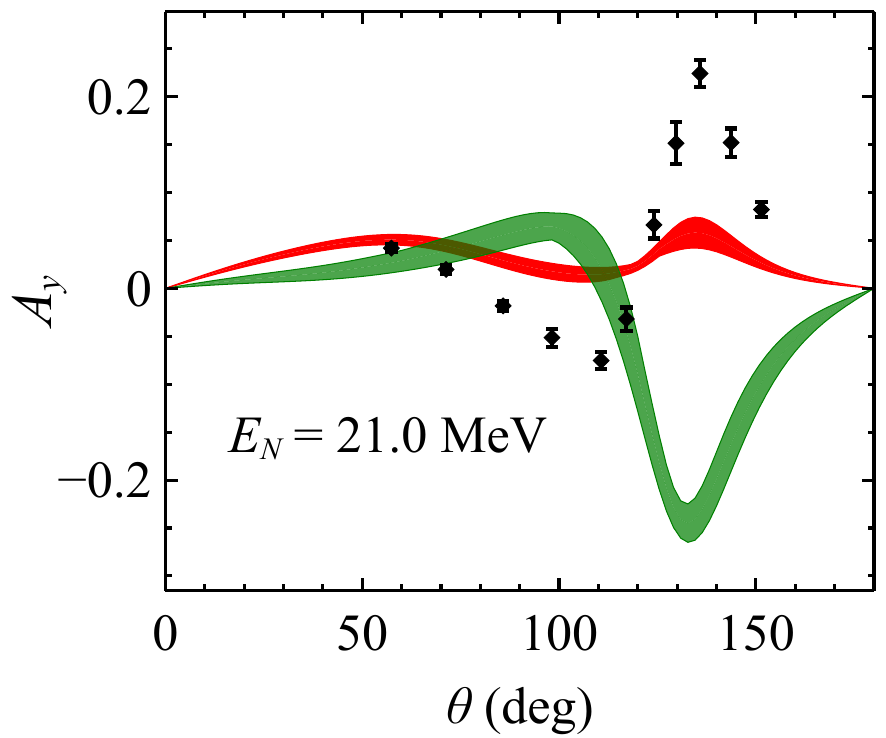}
        \includegraphics[scale=0.28]{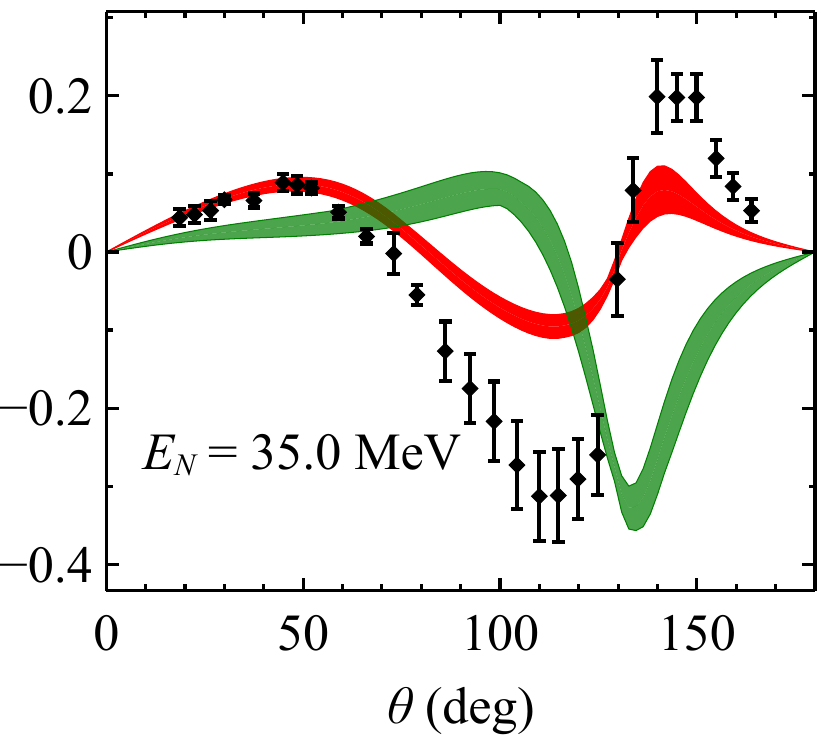}
        \includegraphics[scale=0.28]{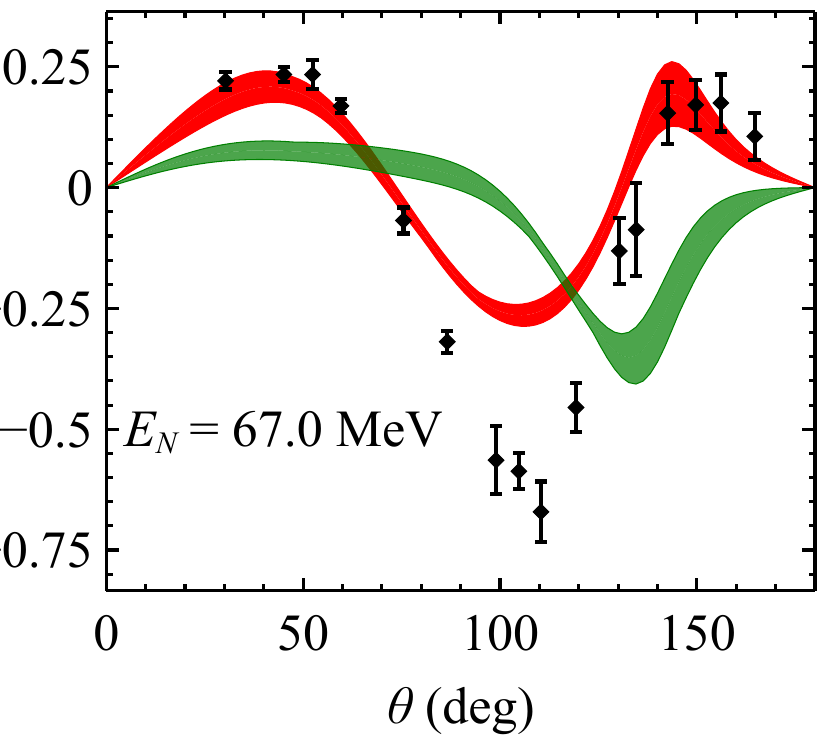}
        \caption{LO and NLO nucleon analyzing powers at various $E_N$ are shown. The bands are generated by varying $\Lambda$ from 400 to 1600 MeV. The $nd$ scattering data at $E_N=3.0$, 10.0 and 14.1, 21.0, and 67.0 MeV are taken from Refs.~\cite{McAninch:1993fzh}, \cite{Howell:1987} (for 10.0 and 14.1 MeV), \cite{Weisel:2015min}, and \cite{Ruhl:1991qop}. The $pd$ data at $E_N=35.0$ MeV are from Ref.~\cite{Bunker:1968omc}.
        }\label{fig:MMW_Ay}
\end{figure*}

The angular distributions of the differential cross sections and the nucleon analyzing power $A_y$ are computed to study how much the NLO corrections change relative to LO.
These observables are computed with the partial-wave sum truncated beyond $J^P=\frac{19}{2}^{\pm}$, at which point including higher waves alters the results by less than $1\%$.
Using $\Lambda = 400-1600$ MeV, Figs.~\ref{fig:MMWDSG} and \ref{fig:MMW_Ay} compare the LO and NLO EFT predictions with available experimental data for these observables at their respective energies.

\begin{figure}
    \centering
    \includegraphics[scale=0.28]{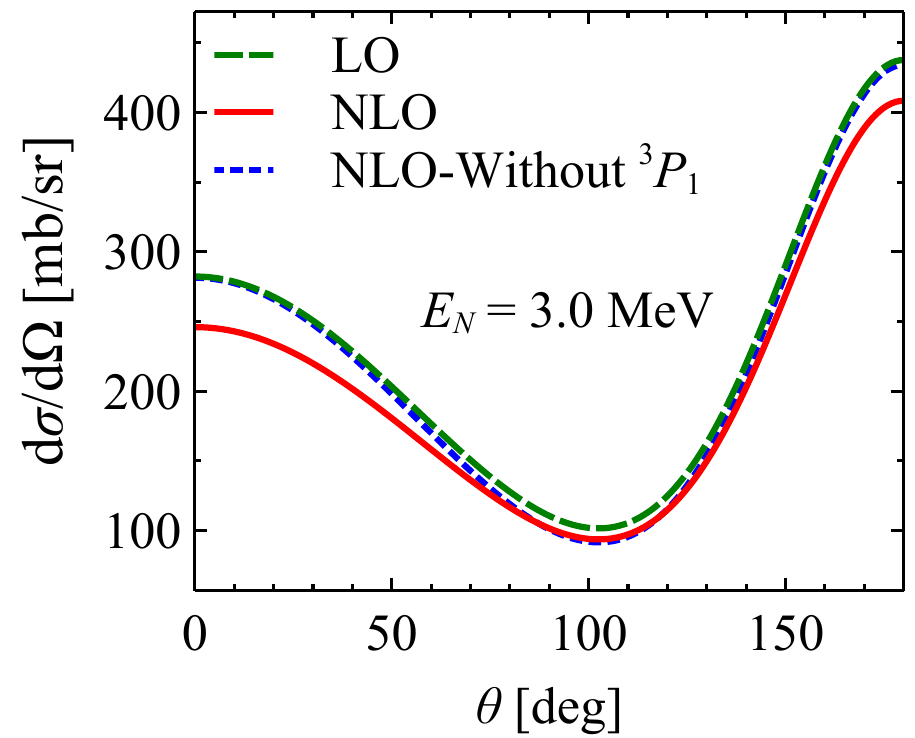}
    \includegraphics[scale=0.28]{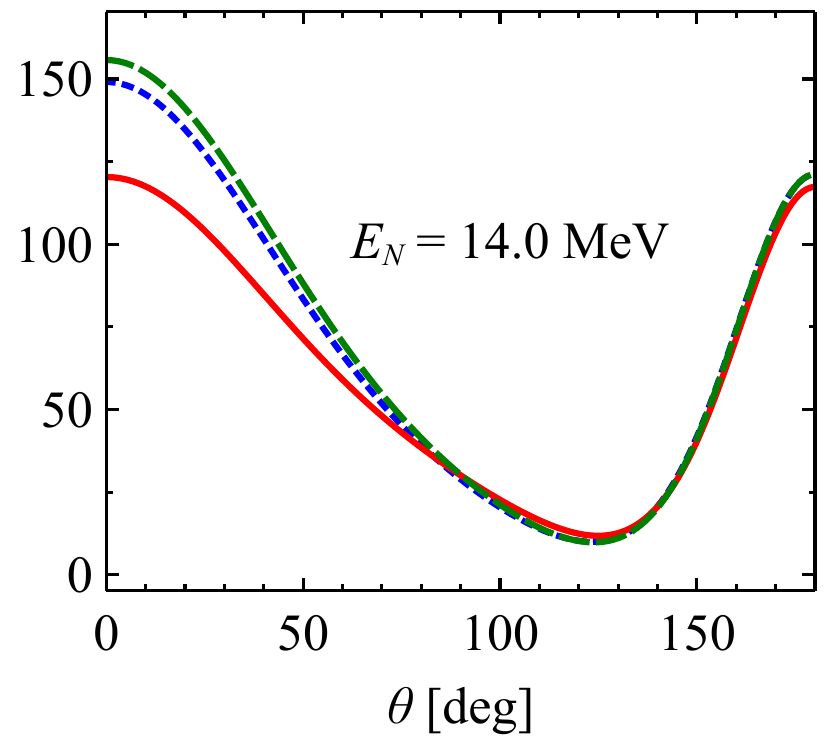}
    \includegraphics[scale=0.28]{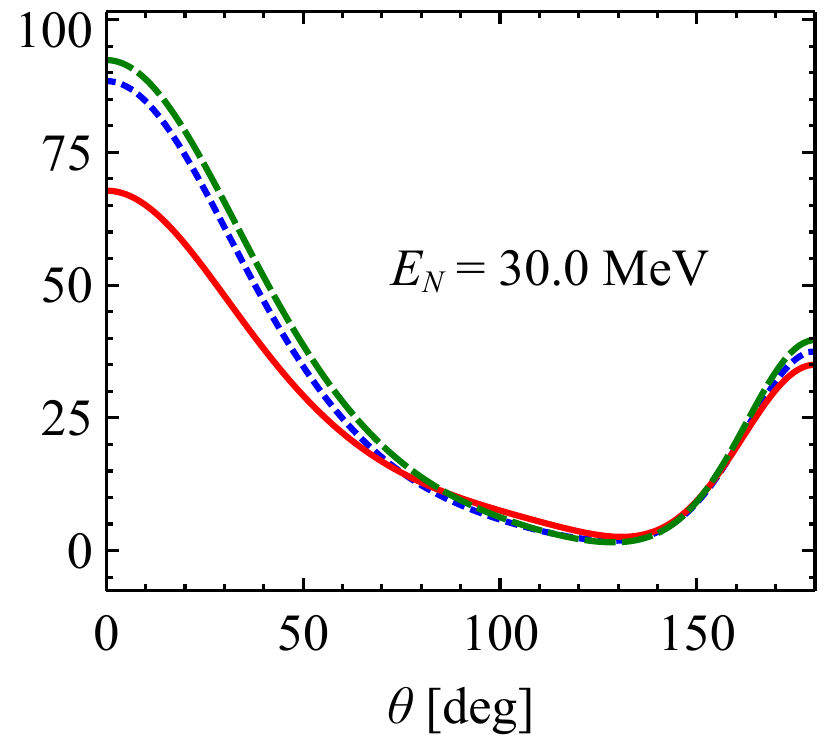}
    \caption{The LO and NLO angular distributions, obtained using the chiral $NN$ potential with $\Lambda=800$ MeV, are shown for various $E_N$. The dotted curves show the NLO results with the $\chp{3}{1}$ OPE turned off.
    }
    \label{fig:MMW_DSGWithout3p1}
\end{figure}

For the angular distribution, we find that the LO results agree with the data better than the NLO ones at forward angles at most energies studied.
We first note that the NLO correction is mostly driven by repulsion in the $\chp{3}{1}$ channel, which consists of only the OPE potential at the NLO.
As shown in Fig.~\ref{fig:MMW_DSGWithout3p1}, removing the $\chp{3}{1}$ component from the NLO potentials results in the angular distributions nearly identical to the LO results.
Because the tensor operator $S_{12}$ has its largest matrix element in $\chp{3}{1}$, compared with other perturbative channels, the OPE tensor force in $\chp{3}{1}$ appears to be the strongest at NLO.
However, it is not entirely clear to us whether other mechanisms exist for the $\chp{3}{1}$ OPE to dominate the NLO correction to the differential cross section.
Using the nucleon-delta mass splitting $\delta = 293$ MeV as the breakdown scale for our delta-less chiral forces, we estimate the EFT expansion error by powers of $\mathrm{max}(q_0, \gamma_d, m_\pi)/\delta$.
For instance, the EFT error for NLO is $(m_{\pi}/\delta)^2\simeq 22\%$ at $E_N = 9.0$ MeV, which is able to explain the discrepancy between the EFT predictions and the data.

The LO, however, agrees worse than the NLO with data in terms of describing $A_y$.
In Fig.~\ref{fig:MMW_Ay}, the maximum of $A_y$ at LO has the opposite sign relative the experimental data. Due to the lack of $nd$ experimental data, we use proton-deuteron ($pd$) data in Fig~.\ref{fig:MMW_Ay} at $E_N=35$ MeV.
The NLO corrections reverse the wrong trend and move toward the data.
This suggests that for the power counting we adopt, $A_y$ can only be described well at quite high orders.

In Ref.~\cite{Zhai:2025grs}, the EKM chiral potentials~\cite{Epelbaum:2014efa} are used to calculate $Nd$ scattering, and the subleading potentials are treated nonperturbatively.
Note that ``NLO'' in the EKM scheme refers to $\mathcal{O}(Q^2)$ corrections to LO and therefore includes more physics than our NLO does, e.g., two-pion-exchange potentials.
By contrast, the perturbative calculations in this work allow us to access a wider cutoff window to test RG invariance, as in the $NN$ sector.
The LO+NLO potentials used here are somewhat comparable to the EKM LO; both have OPE as the long-range force.
In this work, however, OPE is nonperturbative only in $\cs{1}{0}$, $\csd$, and $\chp{3}{0}$, and the setup of the contact interactions differs: our scheme includes a $\chp{3}{0}$ term at LO and a momentum-dependent $\cs{1}{0}$ term at NLO.
While the NLO of this work and the EKM LO yield similar angular distributions, they differ in their description of the analyzing power: our NLO exhibits gradual improvements toward the experimental data.
It will be worthwhile in future work to identify which factor drives this difference: the perturbative treatment of OPE in higher partial waves or the arrangement of contact interactions.

\section{Summary\label{sec:summary}}

We have performed calculations of $Nd$ elastic scattering using renormalization-group-invariant chiral nuclear forces up to next-to-leading order, as developed in Refs.~\cite{Long:2011xw, Long:2012ve, Wu:2018lai}. 
In our implementation of the Faddeev equation, a contour-deformation technique is employed, avoiding subtractions of the singularities of the integral equation.
Strict perturbative calculations for the NLO potential after treating the LO nonperturbatively are performed.
We demonstrated that computational costs can be substantially reduced in perturbation theory by leveraging the fact that the LO potentials are restricted to just a few $NN$ partial waves. 
This approach yields a hierarchy of linear equations characterized by a fixed kernel --- determined solely by the LO $NN$ channels and their matrix elements --- with driving terms that incorporate the higher-order potentials.
In general, the computation time and memory usage of FKPT are at least an order of magnitude smaller than those of the auxiliary-potential method.
The WPCD method was first used to benchmark our techniques. Good agreement was found for the phase shifts, elastic scattering differential cross sections, and nucleon analyzing powers.

We used the FKPT method to calculate the $Nd$ scattering phase shifts for the doublet and quartet $S$ waves.
The phase shifts were found to converge with respect to the cutoff $\Lambda$ up to 1600 MeV.
This finding is consistent with earlier work, where RG invariance of the triton binding energy and $nd$ scattering lengths indicated that three-nucleon forces are not required for renormalization purposes up to NLO.
In comparisons with experimental data, we found that although the NLO calculation underpredicts the differential cross section at forward angles relative to LO, it generally yields better agreement for the analyzing power $A_y$.

In anticipation of studying $Nd$ scattering at next-to-next-to leading order ({\NNLO}), we note that, with a similar power counting, the triton binding energy exhibits severe cutoff dependence at this order near the so-called exceptional cutoffs~\cite{Thim:2025vhe}.
These cutoffs were first reported in Ref.~\cite{Gasparyan:2022isg} in the context of $NN$ scattering and the issue was discussed further in Refs.~\cite{Peng:2024aiz, Yang:2024yqv, Peng:2025ykg, PavonValderrama:2025zzk, PavonValderrama:2025azr}.
Because $Nd$ scattering will allow us to access more channels of the $3N$ system than the triton, we expect the techniques presented here to help us exam whether such cutoff variations persist in other $3N$ channels with different quantum numbers.

\section*{Acknowledgments}

We would like to thank Zeyuan Ye, Kaifei Ning, and Jinlong Dang for their contributions in the early stages of developing the Faddeev code used in this paper.
This work was supported by the National Natural Science Foundation of China (NSFC) under Grant Nos. 12275185, 12335002, and U2067205, and by the National Key R\&D Program of China (No. 2023YFA1606702).
DYP thanks O.A. Rubtsova for providing the codes for solving the Faddeev equations using the WPCD method.

\bibliographystyle{unsrt}
\bibliography{NdscatRefs.bib}

@article{Epelbaum:2014efa,
    author = "Epelbaum, E. and Krebs, H. and Mei{\ss}ner, U. G.",
    title = "{Improved chiral nucleon-nucleon potential up to next-to-next-to-next-to-leading order}",
    eprint = "1412.0142",
    archivePrefix = "arXiv",
    primaryClass = "nucl-th",
    doi = "10.1140/epja/i2015-15053-8",
    journal = "Eur. Phys. J. A",
    volume = "51",
    number = "5",
    pages = "53",
    year = "2015"
}

@article{PavonValderrama:2025azr,
    author = "Pavon Valderrama, Manuel",
    title = "{Reexamining the perturbative renormalizability of coupled triplets}",
    eprint = "2510.15789",
    archivePrefix = "arXiv",
    primaryClass = "nucl-th",
    doi = "10.1103/32q7-8j8r",
    journal = "Phys. Rev. C",
    volume = "113",
    number = "1",
    pages = "014001",
    year = "2026"
}

@article{PavonValderrama:2025zzk,
    author = "Pavon Valderrama, Manuel",
    title = "{Regulator constraints for the perturbative renormalizability of attractive triplets}",
    eprint = "2509.23855",
    archivePrefix = "arXiv",
    primaryClass = "nucl-th",
    doi = "10.1103/2zgt-znp9",
    journal = "Phys. Rev. C",
    volume = "112",
    number = "6",
    pages = "064009",
    year = "2025"
}

@article{Yang:2024yqv,
    author = "Yang, C. -J.",
    title = "{Further theoretical study on the renormalization group aspect of perturbative corrections}",
    eprint = "2410.08845",
    archivePrefix = "arXiv",
    primaryClass = "nucl-th",
    doi = "10.1103/fxx6-2rng",
    journal = "Phys. Rev. C",
    volume = "112",
    number = "1",
    pages = "014004",
    year = "2025"
}

@article{Peng:2025ykg,
    author = "Peng, Rui and Long, Bingwei and Xu, Fu-Rong",
    title = "{Perturbative renormalization of chiral nuclear forces at subleading order in the 3S1{\ensuremath{-}}3D1 channel}",
    eprint = "2508.06838",
    archivePrefix = "arXiv",
    primaryClass = "nucl-th",
    doi = "10.1103/7jjf-gy3m",
    journal = "Phys. Rev. C",
    volume = "112",
    number = "6",
    pages = "064004",
    year = "2025"
}

@article{Peng:2024aiz,
    author = "Peng, Rui and Long, Bingwei and Xu, Fu-Rong",
    title = "{Contact operators in renormalization of attractive singular potentials}",
    eprint = "2407.08342",
    archivePrefix = "arXiv",
    primaryClass = "nucl-th",
    doi = "10.1103/PhysRevC.110.054001",
    journal = "Phys. Rev. C",
    volume = "110",
    number = "5",
    pages = "054001",
    year = "2024"
}

@article{Gasparyan:2022isg,
    author = "Gasparyan, A. M. and Epelbaum, E.",
    title = "{{\textquotedblleft}Renormalization-group-invariant effective field theory{\textquotedblright} for few-nucleon systems is cutoff dependent}",
    eprint = "2210.16225",
    archivePrefix = "arXiv",
    primaryClass = "nucl-th",
    doi = "10.1103/PhysRevC.107.034001",
    journal = "Phys. Rev. C",
    volume = "107",
    number = "3",
    pages = "034001",
    year = "2023"
}

@article{Hammer:2019poc,
    author = {Hammer, H. -W. and K{\"o}nig, S. and van Kolck, U.},
    title = "{Nuclear effective field theory: status and perspectives}",
    eprint = "1906.12122",
    archivePrefix = "arXiv",
    primaryClass = "nucl-th",
    doi = "10.1103/RevModPhys.92.025004",
    journal = "Rev. Mod. Phys.",
    volume = "92",
    number = "2",
    pages = "025004",
    year = "2020"
}

@article{Weinberg:1992yk,
    author = "Weinberg, Steven",
    title = "{Three body interactions among nucleons and pions}",
    eprint = "hep-ph/9209257",
    archivePrefix = "arXiv",
    reportNumber = "UTTG-11-92",
    doi = "10.1016/0370-2693(92)90099-P",
    journal = "Phys. Lett. B",
    volume = "295",
    pages = "114--121",
    year = "1992"
}

@article{Long:2007vp,
    author = "Long, B. and van Kolck, U.",
    title = "{Renormalization of Singular Potentials and Power Counting}",
    eprint = "0707.4325",
    archivePrefix = "arXiv",
    primaryClass = "quant-ph",
    reportNumber = "UA-ET-07-03, UA-ET-07-03",
    doi = "10.1016/j.aop.2008.01.003",
    journal = "Annals Phys.",
    volume = "323",
    pages = "1304--1323",
    year = "2008"
}

@article{Beane:2000wh,
    author = "Beane, S. R. and Bedaque, Paulo F. and Childress, L. and Kryjevski, A. and McGuire, J. and van Kolck, U.",
    title = "{Singular potentials and limit cycles}",
    eprint = "quant-ph/0010073",
    archivePrefix = "arXiv",
    reportNumber = "NT-UW-00-023, DOE-ER-41132-102-INT-00, RBRC-140, KRL-MAP-271, NT@UW-00-023",
    doi = "10.1103/PhysRevA.64.042103",
    journal = "Phys. Rev. A",
    volume = "64",
    pages = "042103",
    year = "2001"
}

@article{PavonValderrama:2004nb,
    author = "Pavon Valderrama, M. and Ruiz Arriola, E.",
    title = "{Renormalization of NN-scattering with one pion exchange and boundary conditions}",
    eprint = "nucl-th/0405057",
    archivePrefix = "arXiv",
    doi = "10.1103/PhysRevC.70.044006",
    journal = "Phys. Rev. C",
    volume = "70",
    pages = "044006",
    year = "2004"
}

@article{Kaplan:2019znu,
    author = "Kaplan, David B.",
    title = "{Convergence of nuclear effective field theory with perturbative pions}",
    eprint = "1905.07485",
    archivePrefix = "arXiv",
    primaryClass = "nucl-th",
    reportNumber = "INT-PUB-19-015",
    doi = "10.1103/PhysRevC.102.034004",
    journal = "Phys. Rev. C",
    volume = "102",
    number = "3",
    pages = "034004",
    year = "2020"
}

@article{Thim:2025vhe,
    author = {Thim, Oliver and Ekstr{\"o}m, Andreas and Forss{\'e}n, Christian},
    title = "{Perturbative {\ensuremath{\chi}}EFT calculations of the deuteron and triton up to N2LO}",
    eprint = "2510.12207",
    archivePrefix = "arXiv",
    primaryClass = "nucl-th",
    doi = "10.1103/tjld-x141",
    journal = "Phys. Rev. C",
    volume = "112",
    number = "6",
    pages = "064008",
    year = "2025"
}

@article{Shi:2022blm,
    author = "Shi, Wenchao and Peng, Rui and Liu, Tai-Xing and Lyu, Songlin and Long, Bingwei",
    title = "{Perturbative calculations of deuteron form factors}",
    eprint = "2205.02000",
    archivePrefix = "arXiv",
    primaryClass = "nucl-th",
    doi = "10.1103/PhysRevC.106.015505",
    journal = "Phys. Rev. C",
    volume = "106",
    number = "1",
    pages = "015505",
    year = "2022"
}

@article{Liu:2022cfd,
    author = "Liu, Tai-Xing and Peng, Rui and Lyu, Songlin and Long, Bingwei",
    title = "{Renormalization of proton-proton fusion in chiral effective field theory}",
    eprint = "2207.04241",
    archivePrefix = "arXiv",
    primaryClass = "nucl-th",
    doi = "10.1103/PhysRevC.106.055501",
    journal = "Phys. Rev. C",
    volume = "106",
    number = "5",
    pages = "055501",
    year = "2022"
}

@article{Thim:2024yks,
    author = {Thim, Oliver and Ekstr{\"o}m, Andreas and Forss{\'e}n, Christian},
    title = "{Perturbative computations of neutron-proton scattering observables using renormalization-group invariant chiral effective field theory up to N3LO}",
    eprint = "2402.15325",
    archivePrefix = "arXiv",
    primaryClass = "nucl-th",
    doi = "10.1103/PhysRevC.109.064001",
    journal = "Phys. Rev. C",
    volume = "109",
    number = "6",
    pages = "064001",
    year = "2024"
}

@article{Thim:2023fnl,
    author = {Thim, Oliver and May, Eleanor and Ekstr{\"o}m, Andreas and Forss{\'e}n, Christian},
    title = "{Bayesian analysis of chiral effective field theory at leading order in a modified Weinberg power counting approach}",
    eprint = "2302.12624",
    archivePrefix = "arXiv",
    primaryClass = "nucl-th",
    doi = "10.1103/PhysRevC.108.054002",
    journal = "Phys. Rev. C",
    volume = "108",
    number = "5",
    pages = "054002",
    year = "2023"
}

@article{Miller:2021vby,
    author = {Miller, Sean B. S. and Ekstr{\"o}m, Andreas and Forss{\'e}n, Christian},
    title = "{Wave-packet continuum discretisation for nucleon{\textendash}nucleon scattering predictions}",
    eprint = "2106.00454",
    archivePrefix = "arXiv",
    primaryClass = "nucl-th",
    doi = "10.1088/1361-6471/ac3cfd",
    journal = "J. Phys. G",
    volume = "49",
    number = "2",
    pages = "024001",
    year = "2022"
}

@article{Zhai:2025grs,
    author = "Zhai, Qing-Yu and Pang, Dan-Yang and Chen, Wen-Di and Rubtsova, O. A. and Xu, Rui-Rui and Lu, Jun-Xu and Liang, Haozhao and Geng, Li-Sheng",
    title = "{Neutron-deuteron scattering revisited with the EKM chiral nuclear force and the WPCD method}",
    journal = "",
    year = "2025",
    month = "12",
    note = "arXiv: 2512.02475 [nucl-th]"   
}

@article{POMERANTSEV2016121,
title = {Fast GPU-based calculations in few-body quantum scattering},
journal = {Computer Physics Communications},
volume = {204},
pages = {121-131},
year = {2016},
issn = {0010-4655},
doi = {https://doi.org/10.1016/j.cpc.2016.03.018},
url = {https://www.sciencedirect.com/science/article/pii/S0010465516300765},
author = {V.N. Pomerantsev and V.I. Kukulin and O.A. Rubtsova and S.K. Sakhiev}
}

@article{Andis:2025fsg,
    author = {Andis, Andrew J. and Lyu, Songlin and Long, Bingwei and K{\"o}nig, Sebastian},
    title = "{Perturbative EFT calculation of the deuteron longitudinal response function}",
    eprint = "2512.12823",
    archivePrefix = "arXiv",
    primaryClass = "nucl-th",
    journal = "",
    month = "12",
    year = "2025",
    note = "arXiv: 2512.12823 [nucl-th]  "
}

@article{Margaryan:2015rzg,
    author = "Margaryan, Arman and Springer, Roxanne P. and Vanasse, Jared",
    title = "{$nd$ scattering and the A$_y$ puzzle to next-to-next-to-next-to-leading order}",
    eprint = "1512.03774",
    archivePrefix = "arXiv",
    primaryClass = "nucl-th",
    doi = "10.1103/PhysRevC.93.054001",
    journal = "Phys. Rev. C",
    volume = "93",
    number = "5",
    pages = "054001",
    year = "2016"
}

@article{Glockle:1996jg,
    author = "Glockle, Walter and Witala, H. and Huber, D. and Kamada, H. and Golak, J.",
    title = "{The Three nucleon continuum: Achievements, challenges and applications}",
    doi = "10.1016/0370-1573(95)00085-2",
    journal = "Phys. Rept.",
    volume = "274",
    pages = "107--285",
    year = "1996"
}

@article{Weinberg:1990rz,
    author = "Weinberg, Steven",
    title = "{Nuclear forces from chiral Lagrangians}",
    reportNumber = "UTTG-31-90",
    doi = "10.1016/0370-2693(90)90938-3",
    journal = "Phys. Lett. B",
    volume = "251",
    pages = "288--292",
    year = "1990"
}

@article{Weinberg:1991um,
    author = "Weinberg, Steven",
    title = "{Effective chiral Lagrangians for nucleon - pion interactions and nuclear forces}",
    reportNumber = "UTTG-03-91",
    doi = "10.1016/0550-3213(91)90231-L",
    journal = "Nucl. Phys. B",
    volume = "363",
    pages = "3--18",
    year = "1991"
}

@misc{NNonline,
    title = "The NN-OnLine",
    howpublished = "\url{http://nn-online.org}",
}

@article{Stoks:1993tb,
    author = "Stoks, V. G. J. and Klomp, R. A. M. and Rentmeester, M. C. M. and de Swart, J. J.",
    title = "{Partial wave analaysis of all nucleon-nucleon scattering data below 350-MeV}",
    doi = "10.1103/PhysRevC.48.792",
    journal = "Phys. Rev. C",
    volume = "48",
    pages = "792--815",
    year = "1993"
}

@article{Long:2011xw,
    author = "Long, Bingwei and Yang, C. J.",
    title = "{Renormalizing Chiral Nuclear Forces: Triplet Channels}",
    eprint = "1111.3993",
    archivePrefix = "arXiv",
    primaryClass = "nucl-th",
    reportNumber = "JLAB-THY-11-1464, INT-PUB-11-038",
    doi = "10.1103/PhysRevC.85.034002",
    journal = "Phys. Rev. C",
    volume = "85",
    pages = "034002",
    year = "2012"
}

@article{Long:2011qx,
    author = "Long, Bingwei and Yang, C. J.",
    title = "{Renormalizing chiral nuclear forces: a case study of 3P0}",
    eprint = "1108.0985",
    archivePrefix = "arXiv",
    primaryClass = "nucl-th",
    reportNumber = "JLAB-THY-11-1401",
    doi = "10.1103/PhysRevC.84.057001",
    journal = "Phys. Rev. C",
    volume = "84",
    pages = "057001",
    year = "2011"
}

@article{Long:2012ve,
    author = "Long, Bingwei and Yang, C. J.",
    title = "{Short-range nuclear forces in singlet channels}",
    eprint = "1202.4053",
    archivePrefix = "arXiv",
    primaryClass = "nucl-th",
    reportNumber = "JLAB-THY-12-1495, INT-PUB-12-001",
    doi = "10.1103/PhysRevC.86.024001",
    journal = "Phys. Rev. C",
    volume = "86",
    pages = "024001",
    year = "2012"
}

@article{Wu:2018lai,
    author = "Wu, Shaowei and Long, Bingwei",
    title = "{Perturbative $NN$ scattering in chiral effective field theory}",
    eprint = "1807.04407",
    archivePrefix = "arXiv",
    primaryClass = "nucl-th",
    reportNumber = "CTP-SCU/2018002, CTP-SCU-2018002",
    doi = "10.1103/PhysRevC.99.024003",
    journal = "Phys. Rev. C",
    volume = "99",
    number = "2",
    pages = "024003",
    year = "2019"
}

@article{Nogga:2005hy,
    author = "Nogga, A. and Timmermans, R. G. E. and van Kolck, U.",
    title = "{Renormalization of one-pion exchange and power counting}",
    eprint = "nucl-th/0506005",
    archivePrefix = "arXiv",
    reportNumber = "FZJ-IKP-TH-2005-19",
    doi = "10.1103/PhysRevC.72.054006",
    journal = "Phys. Rev. C",
    volume = "72",
    pages = "054006",
    year = "2005"
}

@article{PavonValderrama:2011fcz,
    author = "Pavon Valderrama, M.",
    title = "{Perturbative Renormalizability of Chiral Two Pion Exchange in Nucleon-Nucleon Scattering: P- and D-waves}",
    eprint = "1108.0872",
    archivePrefix = "arXiv",
    primaryClass = "nucl-th",
    doi = "10.1103/PhysRevC.84.064002",
    journal = "Phys. Rev. C",
    volume = "84",
    pages = "064002",
    year = "2011"
}

@article{Birse:2005um,
    author = "Birse, Michael C.",
    title = "{Power counting with one-pion exchange}",
    eprint = "nucl-th/0507077",
    archivePrefix = "arXiv",
    doi = "10.1103/PhysRevC.74.014003",
    journal = "Phys. Rev. C",
    volume = "74",
    pages = "014003",
    year = "2006"
}

@book{Glockle-Fewbody83,
    author = "W. Glöckle",
    title = "{The Quantum Mechanical Few-Body Problem}",
    isbn = "978-3-642-82081-6",
    publisher = "Springer, Berlin, Heidelberg",
    month = "8",
    year = "1983"
}

@article{Huber:1995zza,
  title = {Realistic phase shift and mixing parameters for elastic neutron-deuteron scattering: Comparison of momentum space and configuration space methods},
  author = {H\"uber, D. and Gl\"ockle, W. and Golak, J. and Wita\l{}a, H. and Kamada, H. and Kievsky, A. and Rosati, S. and Viviani, M.},
  journal = {Phys. Rev. C},
  volume = {51},
  issue = {3},
  pages = {1100--1107},
  numpages = {0},
  year = {1995},
  month = {Mar},
  publisher = {American Physical Society},
  doi = {10.1103/PhysRevC.51.1100},
  url = {https://link.aps.org/doi/10.1103/PhysRevC.51.1100}
}

@article{Vanasse:2013sda,
  title = {Fully perturbative calculation of $nd$ scattering to next-to-next-to-leading order},
  author = {Vanasse, Jared},
  journal = {Phys. Rev. C},
  volume = {88},
  issue = {4},
  pages = {044001},
  numpages = {21},
  year = {2013},
  month = {Oct},
  publisher = {American Physical Society},
  doi = {10.1103/PhysRevC.88.044001},
  url = {https://link.aps.org/doi/10.1103/PhysRevC.88.044001}
}

@article{Born-B,
title = {Phase Shifts and Mixing Parameters for Elastic Neutron-Deuteron Scattering Above Breakup Threshold},
journal = {Few-Body Systems},
volume = {19},
number = {175-193},
year = {1995},
issn = {1432-5411},
url = {https://doi.org/10.1007/s006010050025},
doi = {10.1007/s006010050025},
author = {D. Hüber and J. Golak and  H. Witala and  W.Glöckle and H.Kamada},
}

@article{Glockle:1982agg,
    author = {Gl\"ockle, W. and Hasberg, G. and Neghabian, A. R.},
    title = "{Numerical treatment of few body equations in momentum space by the Spline method}",
    doi = "10.1007/BF01417437",
    journal = "Z. Phys. A",
    volume = "305",
    number = "3",
    pages = "217--221",
    year = "1982"
}

@article{Schwarz:1983rlz,
title = {Elastic neutron-deuteron scattering in the energy range from 2.5 MeV to 30 MeV},
journal = {Nuclear Physics A},
volume = {398},
number = {1},
pages = {1-18},
year = {1983},
issn = {0375-9474},
doi = {https://doi.org/10.1016/0375-9474(83)90645-0},
url = {https://www.sciencedirect.com/science/article/pii/0375947483906450},
author = {P. Schwarz and H.O. Klages and P. Doll and B. Haesner and J. Wilczynski and B. Zeitnitz and J. Kecskemeti},
}

@article{McAninch:1993fzh,
    author = {McAninch, J. E. and Haeberli, W. and Wita\l{}a, H. and Gl\"ockle, W. and Golak, J.},
    title = "{Analyzing power in nd elastic scattering at E n lab =3 MeV. Measurement and calculation}",
    doi = "10.1016/0370-2693(93)90185-K",
    journal = "Phys. Lett. B",
    volume = "307",
    pages = "13--19",
    year = "1993"
}

@article{Howell:1987,
    author = "Howell, C. R. and Tornow, W. and Murphy, K. and et al.",
    title = "{Comparisons of vector analyzing-power data and calculations for neutron-deuteron elastic scattering from 10 to 14 MeV}",
    doi = "https://doi.org/10.1007/BF01078989",
    journal = "Few-Body Systems",
    volume = "2",
    pages = "19–32",
    year = "1987"
}

@article{Weisel:2015min,
    author = "Weisel, G. J. and Tornow, W. and Esterline, J. H.",
    title = "{Neutron\textendash{}deuteron analyzing power data at En = 21 MeV and the energy dependence of the three-nucleon analyzing power puzzle}",
    doi = "10.1088/0954-3899/42/8/085106",
    journal = "J. Phys. G",
    volume = "42",
    number = "8",
    pages = "085106",
    year = "2015"
}

@article{Bunker:1968omc,
    author = "Bunker, S. N. and Cameron, J. M. and Carlson, R. F. and Richardson, J. Reginald and Toma\v{s}, P. and Van Oers, W. T. H. and Verba, J. W.",
    title = "{Differential cross sections and polarizations in elastic p-d scattering at medium energies}",
    doi = "10.1016/0375-9474(68)90418-1",
    journal = "Nucl. Phys. A",
    volume = "113",
    pages = "461--480",
    year = "1968"
}

@article{Ruhl:1991qop,
    author = {R{\"u}hl, H. and others},
    title = "{Analyzing power in n +d elastic scattering at 67 MeV}",
    doi = "10.1016/0375-9474(91)90275-B",
    journal = "Nucl. Phys. A",
    volume = "524",
    pages = "377--390",
    year = "1991"
}

@article{Seyler:1969sii,
    author = "Seyler, R. G.",
    title = "{Polarization from scattering polarized spin-${\frac{1}{2}}$ on unpolarized spin-1 particles}",
    doi = "10.1016/0375-9474(69)90352-2",
    journal = "Nucl. Phys. A",
    volume = "124",
    pages = "253--272",
    year = "1969"
}

@article{Moiseyev:1998gjp,
    author = "Moiseyev, Nimrod",
    title = "{Quantum theory of resonances: calculating energies, widths and cross-sections by complex scaling}",
    doi = "10.1016/S0370-1573(98)00002-7",
    journal = "Phys. Rept.",
    volume = "302",
    number = "5-6",
    pages = "212--293",
    year = "1998"
}

@article{Ho:1983lwa,
    author = "Ho, Y. K.",
    title = "{The method of complex coordinate rotation and its applications to atomic collision processes}",
    doi = "10.1016/0370-1573(83)90112-6",
    journal = "Phys. Rept.",
    volume = "99",
    number = "1",
    pages = "1--68",
    year = "1983"
}

@article{Myo:2014ypa,
    author = "Myo, Takayuki and Kikuchi, Yuma and Masui, Hiroshi and Kat{\={o}}, Kiyoshi",
    title = "{Recent development of complex scaling method for many-body resonances and continua in light nuclei}",
    eprint = "1410.4356",
    archivePrefix = "arXiv",
    primaryClass = "nucl-th",
    doi = "10.1016/j.ppnp.2014.08.001",
    journal = "Prog. Part. Nucl. Phys.",
    volume = "79",
    pages = "1--56",
    year = "2014"
}

@article{Myo:2020rni,
    author = "Myo, Takayuki and Kato, Kiyoshi",
    title = "{Complex scaling: Physics of unbound light nuclei and perspective}",
    eprint = "2007.12172",
    archivePrefix = "arXiv",
    primaryClass = "nucl-th",
    doi = "10.1093/ptep/ptaa101",
    journal = "PTEP",
    volume = "2020",
    number = "12",
    pages = "12A101",
    year = "2020"
}

@article{Miller:2022beg,
    author = {Miller, Sean B. S. and Ekstr{\"o}m, Andreas and Hebeler, Kai},
    title = "{Neutron-deuteron scattering cross sections~with chiral NN interactions using wave-packet continuum discretization}",
    eprint = "2201.09600",
    archivePrefix = "arXiv",
    primaryClass = "nucl-th",
    doi = "10.1103/PhysRevC.106.024001",
    journal = "Phys. Rev. C",
    volume = "106",
    number = "2",
    pages = "024001",
    year = "2022"
}

@article{Song:2016ale,
    author = "Song, Young-Ho and Lazauskas, Rimantas and van Kolck, U.",
    title = "{Triton binding energy and neutron-deuteron scattering up to next-to-leading order in chiral effective field theory}",
    eprint = "1612.09090",
    archivePrefix = "arXiv",
    primaryClass = "nucl-th",
    doi = "10.1103/PhysRevC.96.024002",
    journal = "Phys. Rev. C",
    volume = "96",
    number = "2",
    pages = "024002",
    year = "2017",
    note = "[Erratum: Phys.Rev.C 100, 019901 (2019)]"
}

@article{Golak:2014ksa,
    author = "Golak, J. and others",
    title = "{Low-energy neutron-deuteron reactions with N3LO chiral forces}",
    eprint = "1410.0756",
    archivePrefix = "arXiv",
    primaryClass = "nucl-th",
    doi = "10.1140/epja/i2014-14177-7",
    journal = "Eur. Phys. J. A",
    volume = "50",
    pages = "177",
    year = "2014"
}

@article{Witala:2000am,
    author = "Witala, H. and Gloeckle, Walter and Golak, J. and Kamada, H. and Kuros-Zolnierczuk, J. and Nogga, A. and Skibinski, R.",
    title = "{Nd elastic scattering as a tool to probe properties of three nucleon forces}",
    eprint = "nucl-th/0010013",
    archivePrefix = "arXiv",
    doi = "10.1103/PhysRevC.63.024007",
    journal = "Phys. Rev. C",
    volume = "63",
    pages = "024007",
    year = "2001"
}

@article{witala_cornelius_gloeckle_1988, 
    title={Elastic scattering and break-up processes in the n-d system}, volume={3},
    ISSN={0177-7963}, 
    number={3},
    journal={Few-Body Syst.},
    author={Witala, H. and Cornelius, T. and Gloeckle, W.}, 
    year={1988}, 
    pages={123–134}
}

@article{Witala:2013ioa,
    author = "Witala, Henryk and Golak, Jacek and Skibinski, Roman and Topolnicki, Kacper",
    title = "{Calculations of three-nucleon reactions with N$^3$LO chiral forces: achievements and challenges}",
    eprint = "1310.0198",
    archivePrefix = "arXiv",
    primaryClass = "nucl-th",
    doi = "10.1088/0954-3899/41/9/094011",
    journal = "J. Phys. G",
    volume = "41",
    pages = "094011",
    year = "2014"
}

@article{Gloeckle:1990bi,
    author = "Gloeckle, Walter and Cornelius, T. and Witala, H.",
    editor = "Fearing, H. W.",
    title = "{Three nucleon scattering: A test for nuclear dynamics}",
    doi = "10.1016/0375-9474(90)90468-2",
    journal = "Nucl. Phys. A",
    volume = "508",
    pages = "115C--130C",
    year = "1990"
}

@article{Witala:1999sg,
    author = "Witala, H. and Kamada, H. and Nogga, A. and Gloeckle, Walter and Elster, C. and Huber, D.",
    title = "{Modern N N force predictions for the total N/D cross-section up to 300-MeV}",
    eprint = "nucl-th/9901047",
    archivePrefix = "arXiv",
    doi = "10.1103/PhysRevC.59.3035",
    journal = "Phys. Rev. C",
    volume = "59",
    pages = "3035--3046",
    year = "1999"
}

@article{Kievsky:1997bd,
    author = "Kievsky, A. and Viviani, M. and Rosati, S.",
    title = "{N - d scattering above the deuteron breakup threshold}",
    eprint = "nucl-th/9706064",
    archivePrefix = "arXiv",
    reportNumber = "IFUP-TH-21-97",
    doi = "10.1103/PhysRevC.56.2987",
    journal = "Phys. Rev. C",
    volume = "56",
    pages = "2987--2991",
    year = "1997"
}

@article{Witaa:2003en,
    author = "Witala, H. and Nogga, A. and Kamada, H. and Gloeckle, Walter and Golak, J. and Skibinski, R.",
    title = "{Modern nuclear force predictions for the neutron deuteron scattering lengths}",
    doi = "10.1103/PhysRevC.68.034002",
    journal = "Phys. Rev. C",
    volume = "68",
    pages = "034002",
    year = "2003"
}

@article{Epelbaum:2019zqc,
    author = "Epelbaum, E. and others",
    title = "{Towards high-order calculations of three-nucleon scattering in chiral effective field theory}",
    eprint = "1907.03608",
    archivePrefix = "arXiv",
    primaryClass = "nucl-th",
    doi = "10.1140/epja/s10050-020-00102-2",
    journal = "Eur. Phys. J. A",
    volume = "56",
    number = "3",
    pages = "92",
    year = "2020"
}

@article{Witala:2022rzl,
    author = "Wita{\l}a, H. and Golak, J. and Skibi{\'n}ski, R.",
    title = "{Significance of chiral three-nucleon force contact terms for understanding of elastic nucleon-deuteron scattering}",
    eprint = "2203.08499",
    archivePrefix = "arXiv",
    primaryClass = "nucl-th",
    doi = "10.1103/PhysRevC.105.054004",
    journal = "Phys. Rev. C",
    volume = "105",
    number = "5",
    pages = "054004",
    year = "2022"
}

@article{Girlanda:2023znc,
    author = "Girlanda, L. and Filandri, E. and Kievsky, A. and Marcucci, L. E. and Viviani, M.",
    title = "{Effect of the N3LO three-nucleon contact interaction on p-d scattering observables}",
    eprint = "2302.03468",
    archivePrefix = "arXiv",
    primaryClass = "nucl-th",
    doi = "10.1103/PhysRevC.107.L061001",
    journal = "Phys. Rev. C",
    volume = "107",
    number = "6",
    pages = "L061001",
    year = "2023"
}

@article{Konig:2019xxk,
    author = {K{\"o}nig, Sebastian},
    title = "{Energies and radii of light nuclei around unitarity}",
    eprint = "1910.12627",
    archivePrefix = "arXiv",
    primaryClass = "nucl-th",
    doi = "10.1140/epja/s10050-020-00098-9",
    journal = "Eur. Phys. J. A",
    volume = "56",
    number = "4",
    pages = "113",
    year = "2020"
}

@article{Hetherington:1965zza,
    author = "Hetherington, J. H. and Schick, L. H.",
    title = "{Exact Multiple-Scattering Analysis of Low-Energy Elastic K--d Scattering with Separable Potentials}",
    doi = "10.1103/PhysRev.137.B935",
    journal = "Phys. Rev.",
    volume = "137",
    pages = "B935--B948",
    year = "1965"
}

@article{Aaron:1966zz,
    author = "Aaron, R. and Amado, R. D.",
    title = "{Theory of the Reaction n+d --{\ensuremath{>}} n+n+p}",
    doi = "10.1103/PhysRev.150.857",
    journal = "Phys. Rev.",
    volume = "150",
    pages = "857--866",
    year = "1966"
}

\end{document}